\documentclass[reprint,amsmath,amssymb,aps,floatfix]{revtex4-2}
\usepackage[utf8]{inputenc} % umlaut input 
\usepackage[T1]{fontenc}    % umlaut encoding
\usepackage[svgnames,psnames]{xcolor}
\usepackage[colorlinks,citecolor=DarkGreen,linkcolor=FireBrick,linktocpage,unicode]{hyperref}
\usepackage{graphicx,url,qrcode,bm,tikz,dsfont,slashed}

\usetikzlibrary{calc}
\usepackage{cleveref}
\usepackage{here}
\crefname{section}{Sec.}{Secs.}
\Crefname{section}{Section}{Sections}
\crefname{figure}{Fig.}{Figs.}
\crefname{equation}{Eq.}{Eqs.}
\crefname{appendix}{Appendix}{Appendices}
\newcommand{\red}{\color{black}}

\renewcommand{\vec}[1]{\bm{#1}}

\usepackage[normalem]{ulem}
\usepackage[arrowdel]{physics}

\newcommand{\MeV}{\mathrm{\ MeV}}
\newcommand{\fm}{\mathrm{\ fm}}

\def\fm{\text{ fm}}

\def\Julich{J\"{u}lich '04}

\begin{document}

\preprint{APS/123-QED}

\title{$K^- d\rightarrow \pi \Lambda N$ reaction for studying charge symmetry breaking in the $\Lambda N$ interaction}
% Force line breaks with \\

\author{Yutaro Iizawa}

\email{iizawa.y.aa@m.titech.ac.jp}
\affiliation{Department of Physics, Tokyo Institute of Technology, 2-12-1 Ookayama, Megro, Tokyo 152-8551, Japan}

\author{Daisuke Jido}
\affiliation{Department of Physics, Tokyo Institute of Technology, 2-12-1 Ookayama, Megro, Tokyo 152-8551, Japan}

\author{Takatsugu Ishikawa}
\affiliation{Research Center for Electron Photon Science (ELPH),
    Tohoku University, Sendai 982-0826, Japan}

\date{\today}

\begin{abstract}
    We discuss charge symmetry breaking in the $\Lambda N$ interaction. In order to investigate the $\Lambda p$ and $\Lambda n$ interactions at low energies, we propose to utilize  the $K^-d\to \pi^-\Lambda p$ and $K^-d\to \pi^0\Lambda n$ reactions. They are symmetric under the exchange of both the pion and nucleon isospin partners in the final states. This advantage allows us to study charge symmetry breaking in the $\Lambda N$ interaction. We calculate the differential cross sections of these reactions with stopped kaons theoretically and discuss the possibility to extract the scattering length and effective range of the $\Lambda N$ scattering. With stopped kaons,
    % the spin triplet $\Lambda N$ interaction can be extracted 
    the $\Lambda N$ interaction takes place {\red dominantly} in the spin-triplet state
    thanks to the deuteron spin and $s$-wave dominance of the scattering amplitudes at the low energy. We find that
    % the ratio of the $\Lambda N$ invariant mass spectra 
    the ratio of the differential cross sections as a function of the $\Lambda N$ invariant mass
    between the two reactions is useful for revealing how charge symmetry breaking, or isospin symmetry breaking appears in the low-energy $\Lambda N$ scattering.
\end{abstract}

\maketitle

%\tableofcontents

\section{Introduction}
Charge symmetry which is defined as invariance under the rotation by $\pi$ about the $y$ axis in isospin space is one of the fundamental symmetries of strong interaction. It comes from the fact that the electric charges are less relevant for strong interaction and the up and down quark masses are equally small. Thanks to the charge symmetry the properties of nuclei and the nuclear force are very similar when all the protons and neutrons are replaced with neutrons and protons, respectively. When the charge symmetry is extended to hyperon systems, the $\Lambda p$ and $\Lambda n$ scatterings are charge symmetric partners. Charge symmetry breaking is caused by the mass difference of the up and down quarks and electromagnetic effects.
In general, isospin invariance leads to charge independence, which states invariance under rotations in isospin space. In the $\Lambda N$ system, only the weaker charge symmetry appears since isospin of $\Lambda N$ is $1/2$.

Recently a large isospin symmetry breaking effect between $\Lambda p$ and $\Lambda n$ has been suggested by experimental analysis for the $A = 4$ mirror hypernuclei \cite{J-PARCE13:2015uwb,A1:2015isi}. From these experiments data, one finds
% the difference of the excitation energies from the $0^+$ ground state to the $1^-$ excited state of $^4 _\Lambda$H and $^4 _\Lambda$He to be 0.3$\MeV$, which is much larger than the binding energy difference of $^3$H and $^3$He, 0.071$\MeV$. 
the excitation energies of the first $1^-$ states are different by $0.3 \MeV$ between $^4 _\Lambda$H and $^4 _\Lambda$He (both the ground states are $0^+$), while those are different by $0.071 \MeV$ between $^3$H and $^3$He after correcting electromagnetic effects \cite{Miller:1994zh}.
Therefore, one expects a large difference between the $\Lambda p$ and $\Lambda n$ interactions.
The scattering length $a_{\Lambda p}$ and effective range $r_{\Lambda p}$ for $\Lambda p$ scattering were determined experimentally by analyzing the $\Lambda p$ final state interaction in the $p+p\rightarrow K^+ +(\Lambda+p)$ reaction \cite{Budzanowski:2010ib} and their values are $a^s_{\Lambda p} = -2.43^{+0.16}_{-0.25}\fm$ and $r^s_{\Lambda p} = 2.21^{+0.16}_{-0.36}\fm$ for the spin-singlet channel, and $a^t_{\Lambda p} = -1.56^{+0.19}_{-0.22}\fm$ and $r^t_{\Lambda p} = 3.7^{+0.6}_{-0.6}\fm$ for the spin-triplet channel. Here we use the sign convention as positive  (negative) scattering length for repulsive (attractive) interaction. On the other hand, the low-energy $\Lambda n$ scattering parameters have not been experimentally determined yet.

For theoretical approaches, several phenomenological investigations of the $\Lambda N$ interaction have been performed by using boson-exchange models (Nijmegen \cite{Rijken:1998yy,Rijken:2010zzb,Nagels:2015lfa}, J\"{u}lich \cite{Holzenkamp:1989tq,Reuber:1993ip,Haidenbauer:2005zh} and Ehime \cite{Tominaga:1998iy,Tominaga:2001ra}), quark models \cite{Kohno:1999nz,Fujiwara:2006yh,Garcilazo:2007ss}, and hybrid model known as Kyoto-Niigata \cite{Fujiwara:1996qj}. Effective field theory approaches also have investigated the $\Lambda N$ interactions based on the SU(3) chiral symmetry \cite{Savage:1995kv,Korpa:2001au,Polinder:2006zh,Haidenbauer:2013oca,Haidenbauer:2019boi,Li:2016mln,Li:2016paq,Song:2018qqm,Ren:2019qow,Petschauer:2020urh,Haidenbauer:2021wld}. Among them isospin symmetry breaking in the $\Lambda N$ interactions was investigated in Ref.~\cite{Haidenbauer:2021wld}, giving a difference of the $\Lambda N$ scattering lengths $\Delta a^{\rm CSB}\equiv a_{\Lambda p}-a_{\Lambda n}$ of $0.62 \pm 0.08 \fm$ for the spin-singlet state and $-0.10 \pm 0.02 \fm$ for the spin-triplet state.

    {\red We propose the $K^-d \to \pi \Lambda N$ reaction with stopped kaons in order to extract the low-energy $\Lambda N$ properties from the final state interactions.
        One of the good advantages of the reaction is that it includes $\pi^-\Lambda p$ and $\pi^0\Lambda n$ in the final states isospin-symmetrically. Thus we can study isospin breaking by taking the cross section ratio of these two final states. In addition, we can use the same theoretical framework between the two reactions and the reaction calculation can be fixed by the  $\Lambda p$ channel to apply to uncertain $\Lambda n$ channel. The spin of the $\Lambda N$ system is either singlet or triplet. If one considers stopped kaons, the spin-triplet configuration dominates the $\Lambda N$ final state interaction around its production threshold because the deuteron has spin $1$ and the non-spin-flip $s$-wave interactions are the main contributions at low energies.
        This is a good feature to fix the spin configuration of the $\Lambda N$ system. It is also reported in Refs.~\cite{Kotani:1959,Day:1959} that, thanks to the finite size of the deuteron, $K^-$ in the $p$-orbit is also absorbed by the $s$-wave $K^- N$ interaction.

        Historically the $K^- d\to \pi^- \Lambda p$ reaction has been studied to investigate the $\Lambda p$ scattering and also the $\Sigma N$ interaction with kaons at rest \cite{Dahl:1961zzb,Tan:1969jq} and in-flight \cite{Cline:1968,Alexander:1969,Eastwood:1971,Braun:1977ma}.
        There are many theoretical calculations of this reaction with kaons at rest so far \cite{Fujii:1958,Karplus:1959,Kotani:1959,Kudryavtsev:1971,Satoh:1975,Toker:1979pnx,Dalitz:1979qv,Toker:1981zh,Dalitz:1980zc,Dalitz:1982tb,Torres:1986mr,Machner:2013hs}.
        Some of these works have mainly concerned a possible bound state below the $\Sigma N$ threshold by considering $\Sigma$-$\Lambda$ conversion in the intermediate states. Here, focusing on the $\Lambda p$ and $\Lambda n$ interactions, we revisit this reaction by using modern meson-baryon $\bar KN \to MB$ scattering amplitudes obtained by the chiral unitary approach \cite{Oset:2001cn,Jido:2010rx,Jido:2012cy,Yamagata-Sekihara:2012rij} and introducing isospin breaking in the $\Lambda N$ interaction and the scattering amplitudes.}

The structure of this paper is as follows. In Sec.~\ref{formulation}, we explain our theoretical formalism to calculate the $K^- d\rightarrow \pi \Lambda N$ reactions. In Sec.~\ref{results} we show our numerical results and discuss to investigate isospin symmetry breaking in the $\Lambda N$ interaction.
Section ~\ref{sec:Conclusion} is devoted to the summary and conclusion.

\section{Formulation}
\label{formulation}

\subsection{Kinematics}
The reaction $K^-d \rightarrow \pi\Lambda N$ requires five kinematical variables to fix the phase space of the three-body final state \cite{ParticleDataGroup:2020ssz}. In this study we are interested in the mass spectra of the $\Lambda N$ systems, thus we choose the following variables for unpolarized deuteron targets: the $\Lambda N$ invariant mass $M_{\Lambda N}$, the solid angle of the final pion in the total center-of-mass (c.m.) frame $\Omega_\pi$, and the solid angle of the final $\Lambda$ in the $\Lambda$-nucleon c.m.~frame $\Omega^*_{\Lambda}$. Considering stopped kaons,
% we can take the final pion momentum as a reference of the spatial coordinate for the final state without loss of generality. 
a reference of the coordinate is taken along the direction of the final-pion emission.
In addition, once one fixes the reaction plane, the scattering amplitude does not depend on the azimuthal angle of $\Omega^*_{\Lambda}$. The cross section of the reaction is calculated by
\begin{align}
    \label{cross section}
    d \sigma = \frac{1}{ (2\pi)^{3}}
    \frac{M_{d}M_{\Lambda} M_{N}}{2k_{\rm c.m.}E_{\rm c.m.}^{2}}
    \, |{\cal T}|^{2}
    |\vec p_{\pi}|\,
    |\vec p_{\Lambda}^{\, *}|\, dM_{\Lambda N} d\cos\theta^*_{\Lambda}
\end{align}
where $E_\mathrm{c.m.}$ is the total c.m.~energy, $k_\mathrm{c.m.}$ and $\vec p_{\pi}$ are the initial $K^-$ and final $\pi$ momenta in the total c.m. frame, respectively, $\vec p_{\Lambda}^{\, *}$ and $\theta_{\Lambda}^{\, *}$ denote the momentum and the polar angle of the final $\Lambda$ in the $\Lambda$-nucleon c.m. frame, respectively, and $\cal T$ represents the $T$-matrix of the reaction.

\subsection{$K^- d$ scattering amplitudes}
\begin{figure}[htbp]
    \begin{minipage}[b]{0.3\linewidth}
        \centering
        \includegraphics[width=2.5cm,clip]{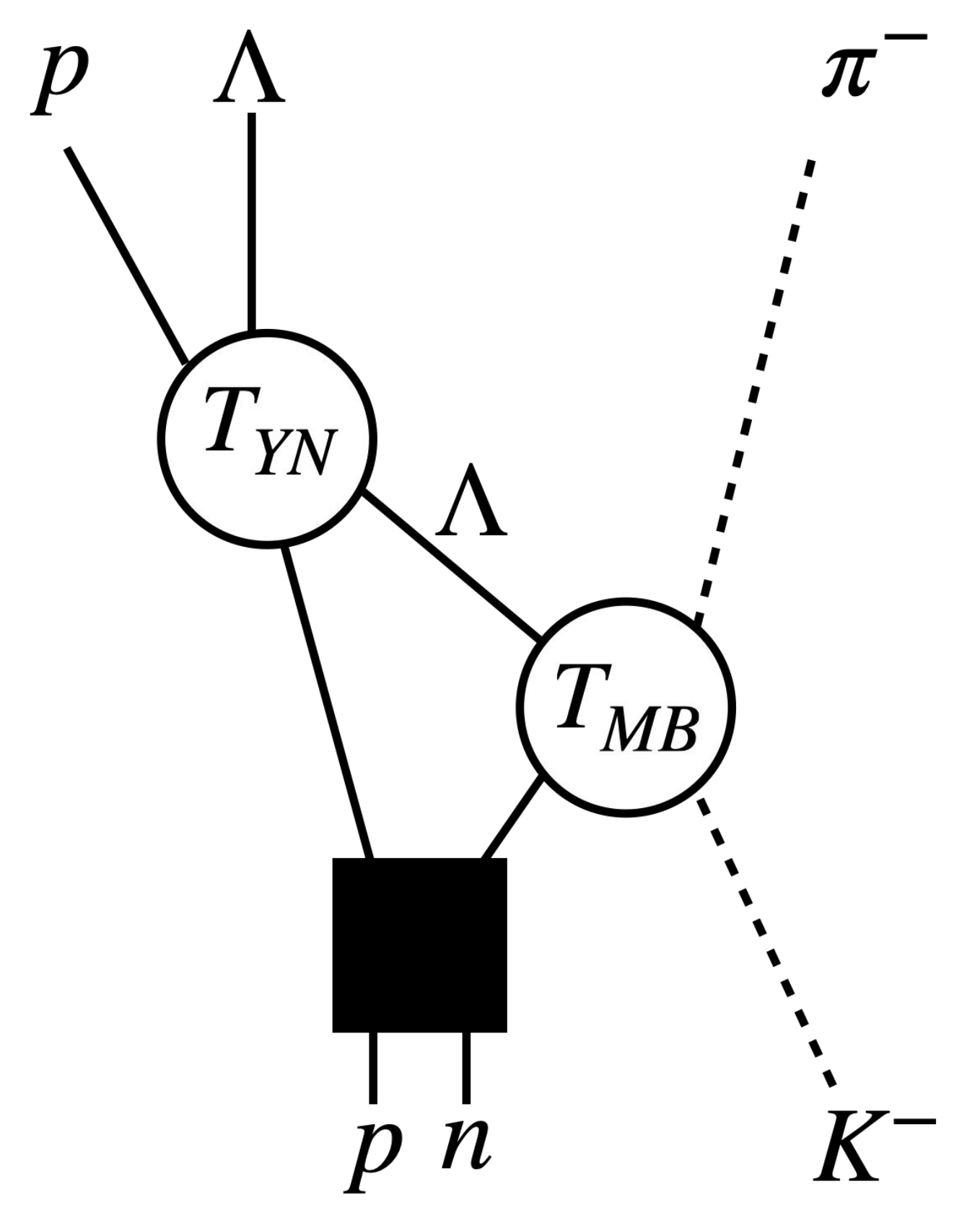}\\
        {\scriptsize Dia.~1}
    \end{minipage}
    \begin{minipage}[b]{0.3\linewidth}
        \centering
        \includegraphics[width=2.5cm,clip]{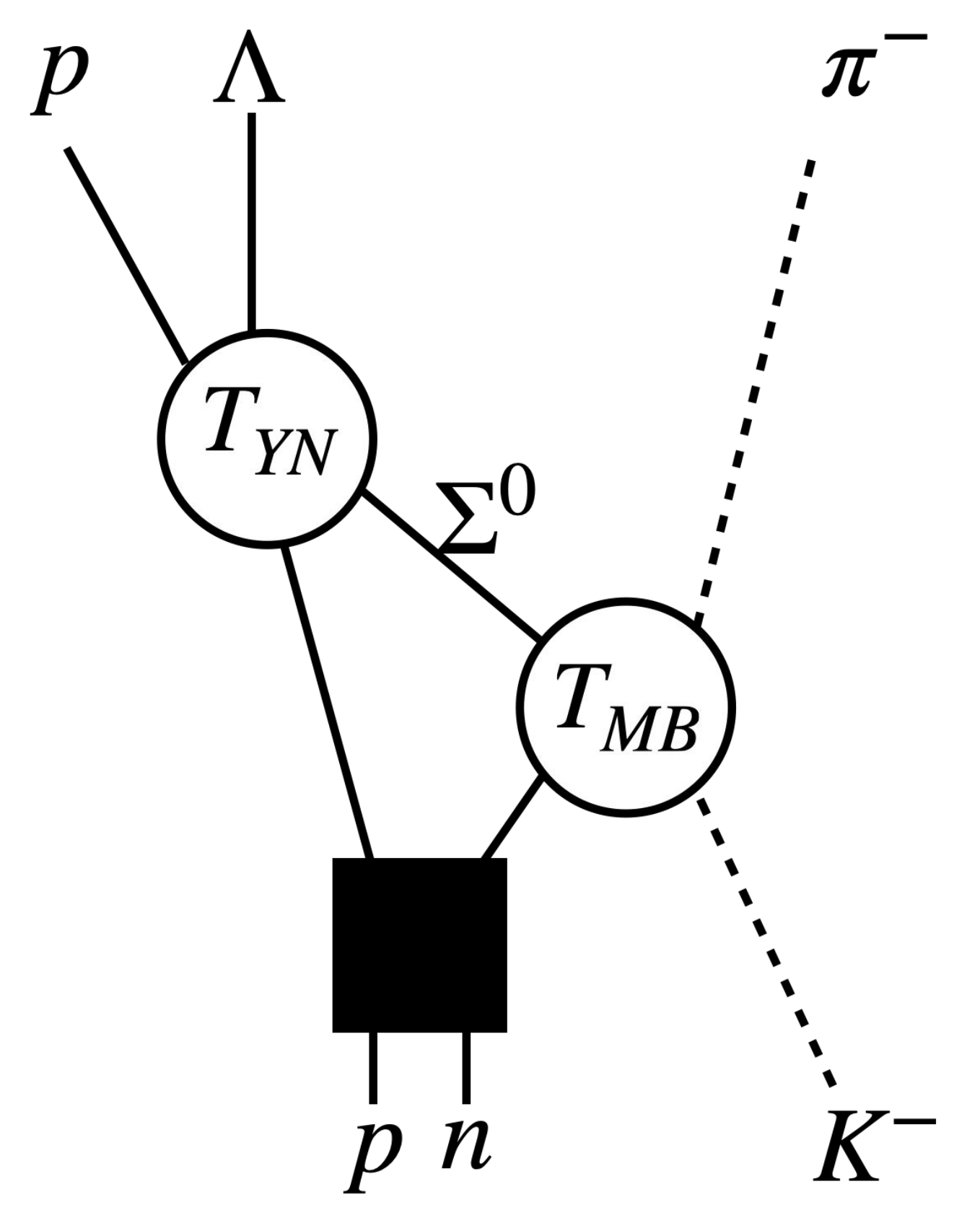}\\
        {\scriptsize Dia.~2a}
    \end{minipage}
    \begin{minipage}[b]{0.3\linewidth}
        \centering
        \includegraphics[width=2.5cm,clip]{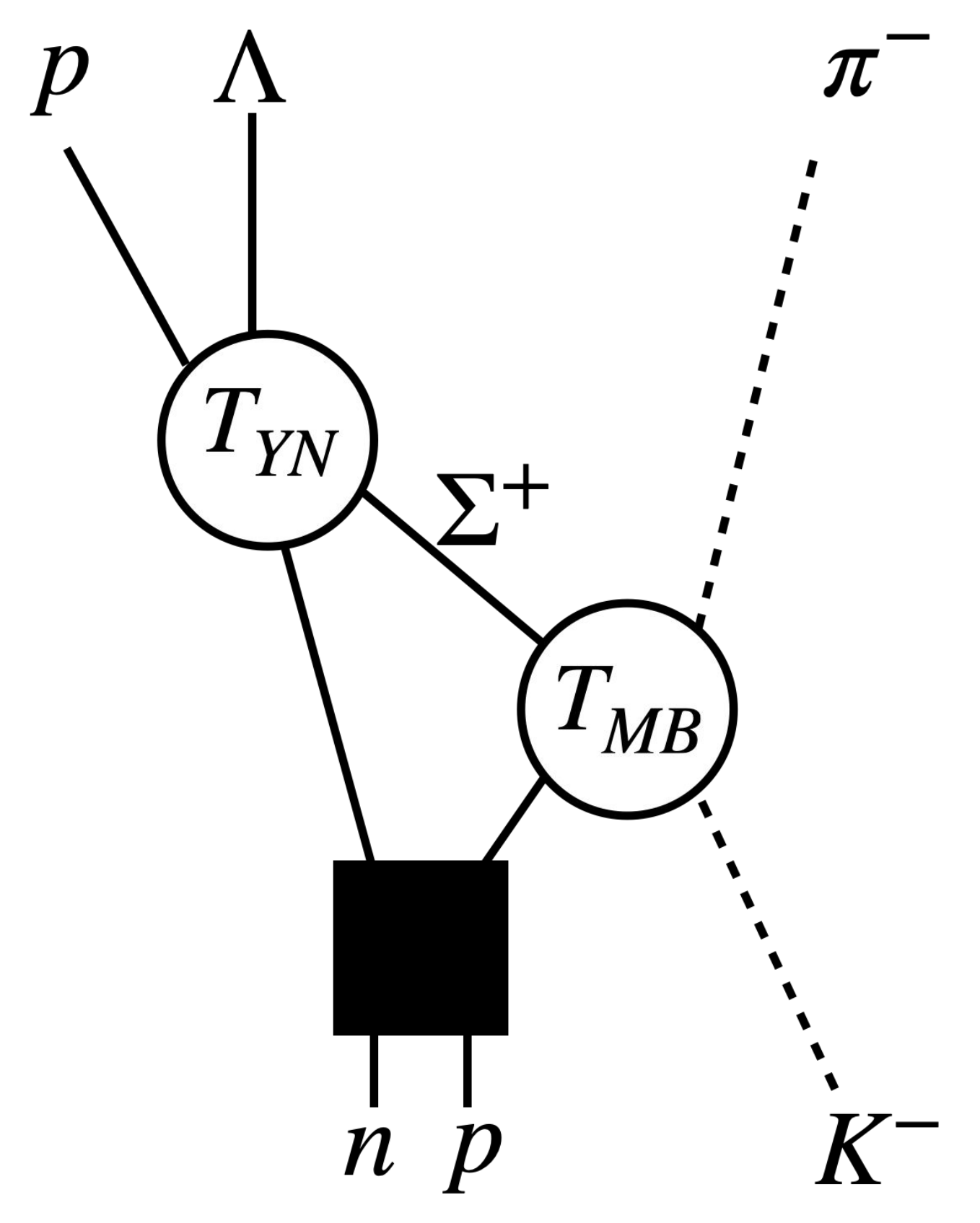}\\
        {\scriptsize Dia.~2b}
    \end{minipage}\\
    \vspace{0.2cm}
    \begin{minipage}[b]{0.3\linewidth}
        \centering
        \includegraphics[width=2.5cm,clip]{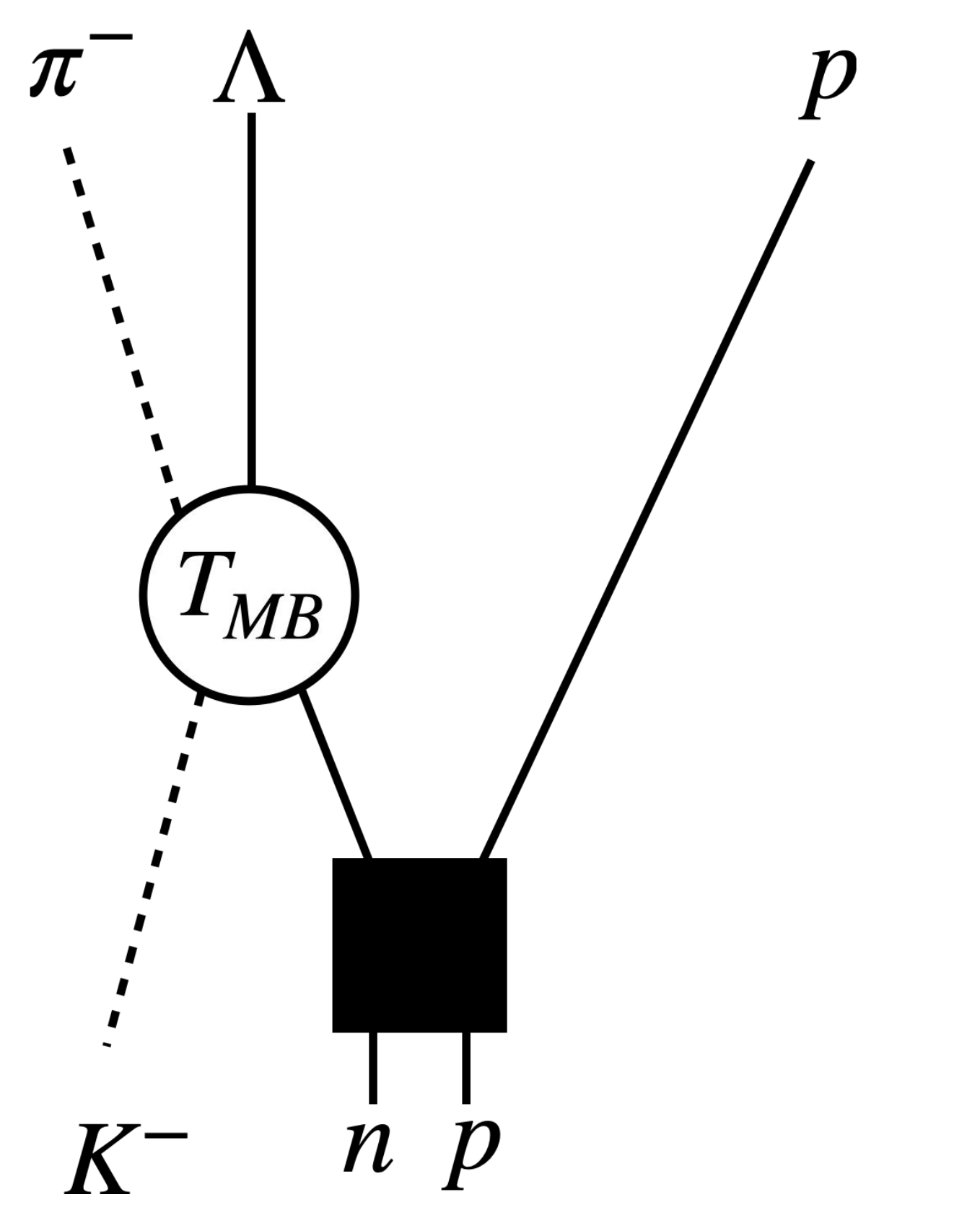}\\
        {\scriptsize Dia.~3}
    \end{minipage}
    \begin{minipage}[b]{0.3\linewidth}
        \centering
        \includegraphics[width=2.5cm,clip]{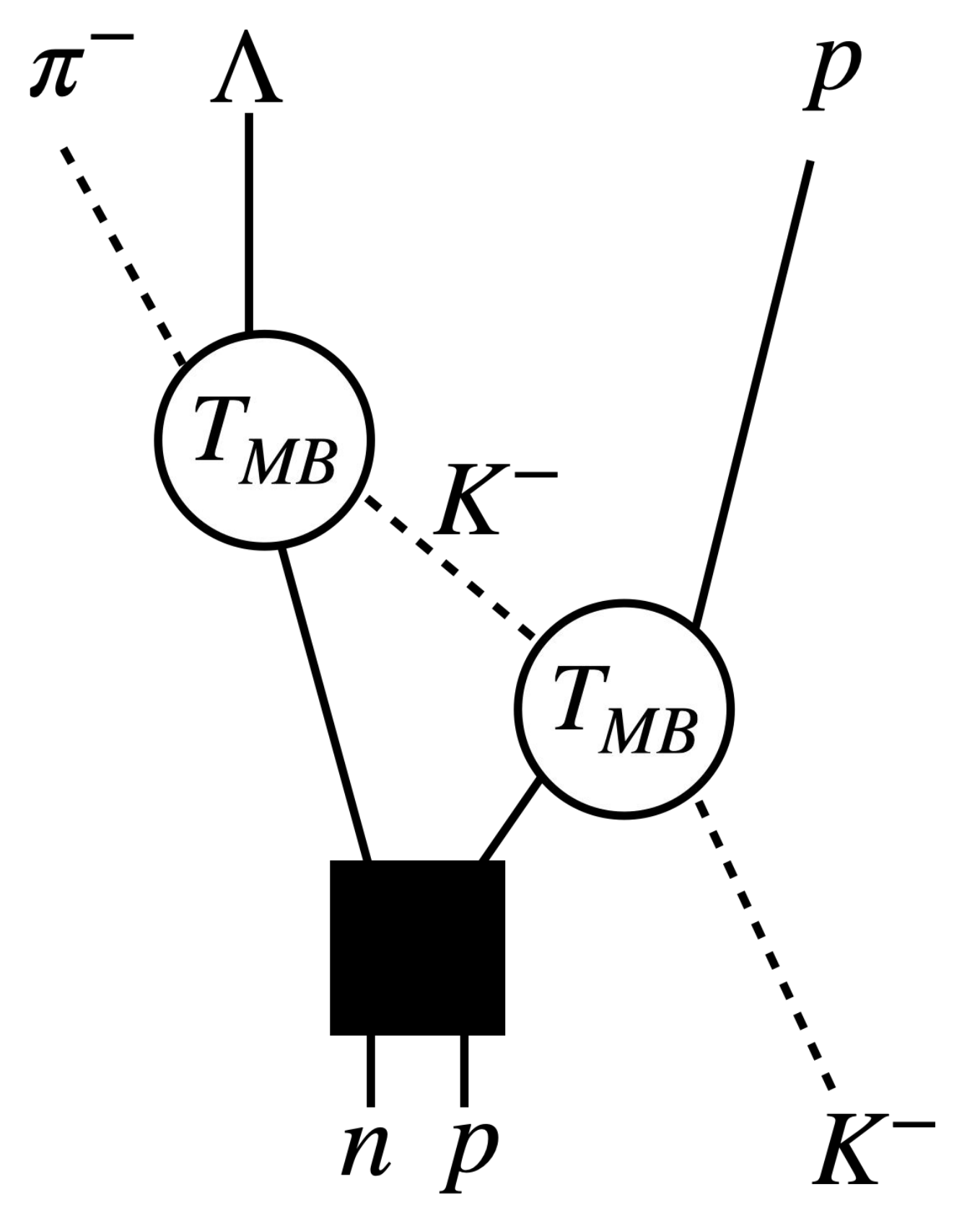}\\
        {\scriptsize Dia.~4}
    \end{minipage}\\
    \vspace{0.2cm}
    \begin{minipage}[b]{0.3\linewidth}
        \centering
        \includegraphics[width=2.5cm,clip]{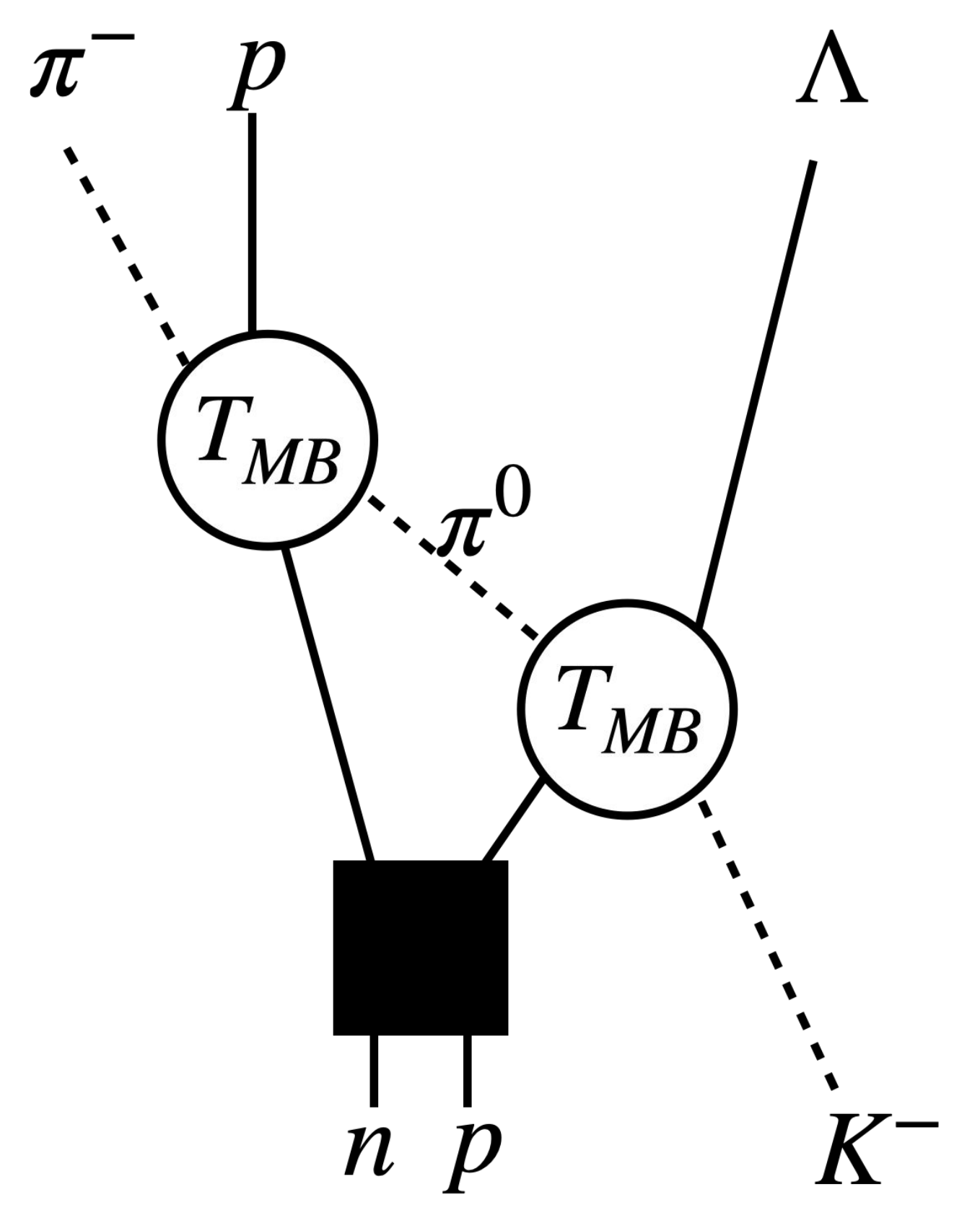}\\
        {\scriptsize Dia.~5a}
    \end{minipage}
    \begin{minipage}[b]{0.3\linewidth}
        \centering
        \includegraphics[width=2.5cm,clip]{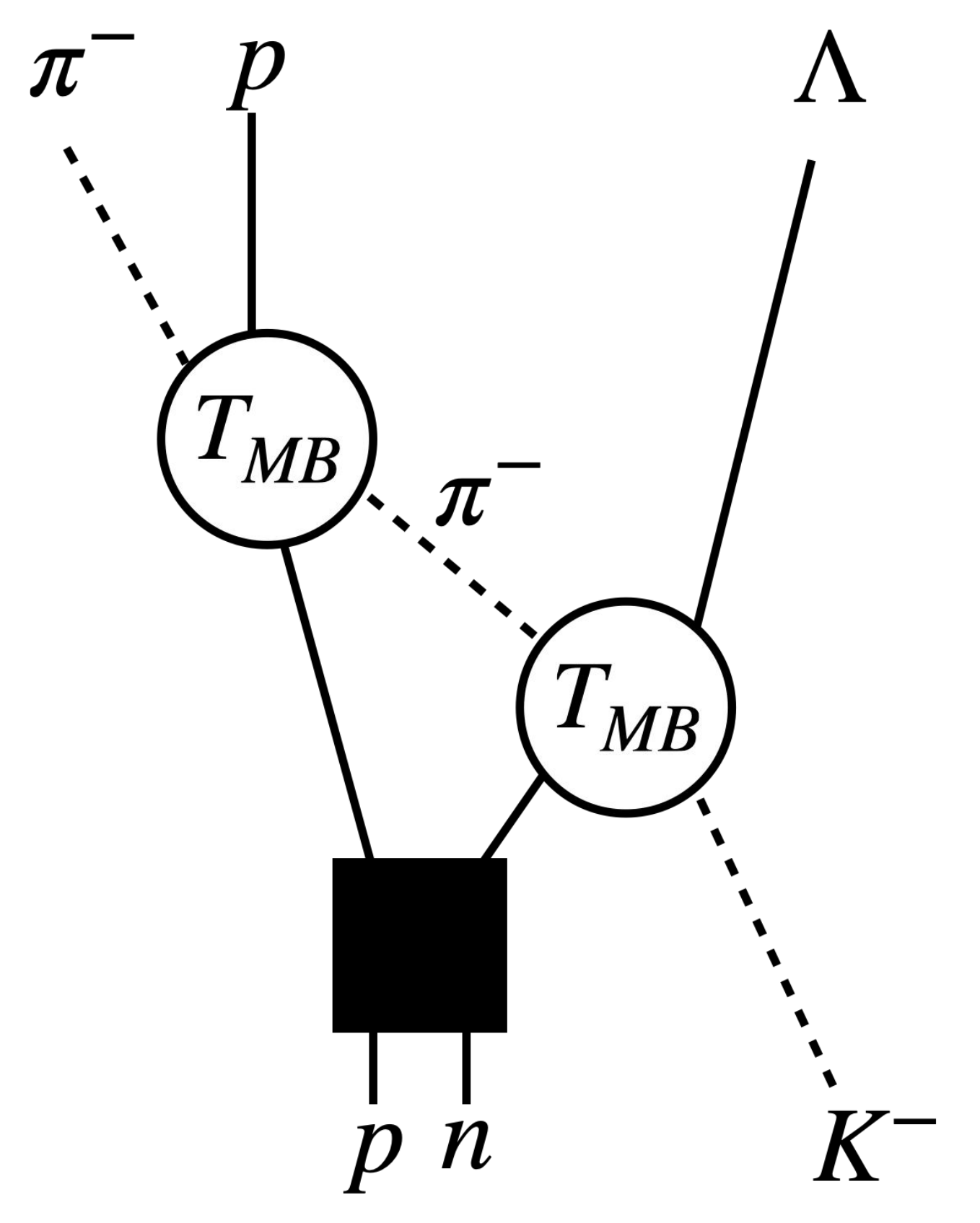}\\
        {\scriptsize Dia.~5b}
    \end{minipage}
    \caption{Feynman diagrams used for the calculation of the $\Lambda p$ process. In the diagrams, $T_{MB}$ ($T_{YN}$) denotes the meson-baryon (hyperon-nucleon) amplitude scattering amplitude.}
    \label{fig:diangram_p}
\end{figure}

\begin{figure}[htbp]
    \begin{minipage}[b]{0.3\linewidth}
        \centering
        \includegraphics[width=2.5cm,clip]{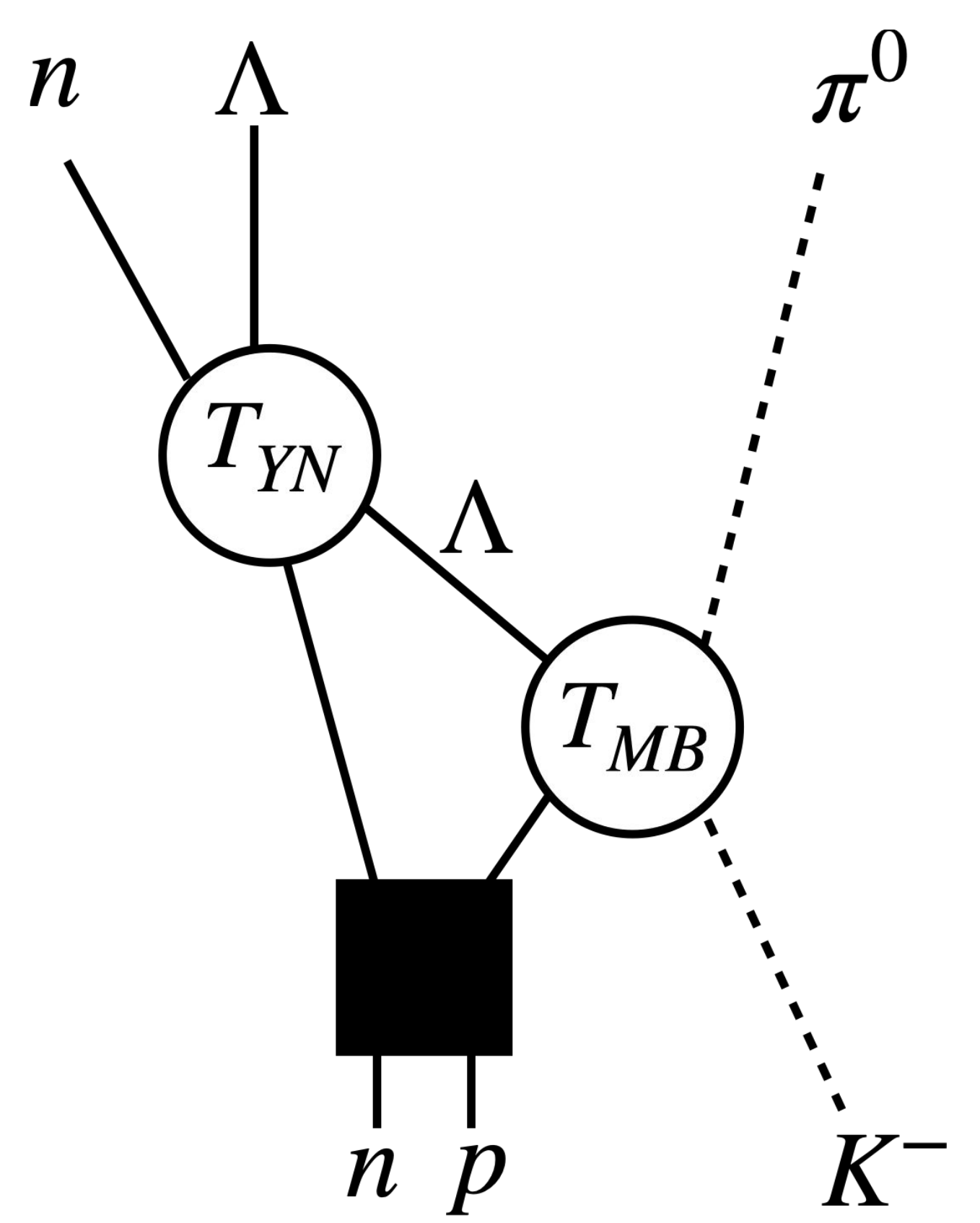}\\
        {\scriptsize Dia.~1}
    \end{minipage}
    \begin{minipage}[b]{0.3\linewidth}
        \centering
        \includegraphics[width=2.5cm,clip]{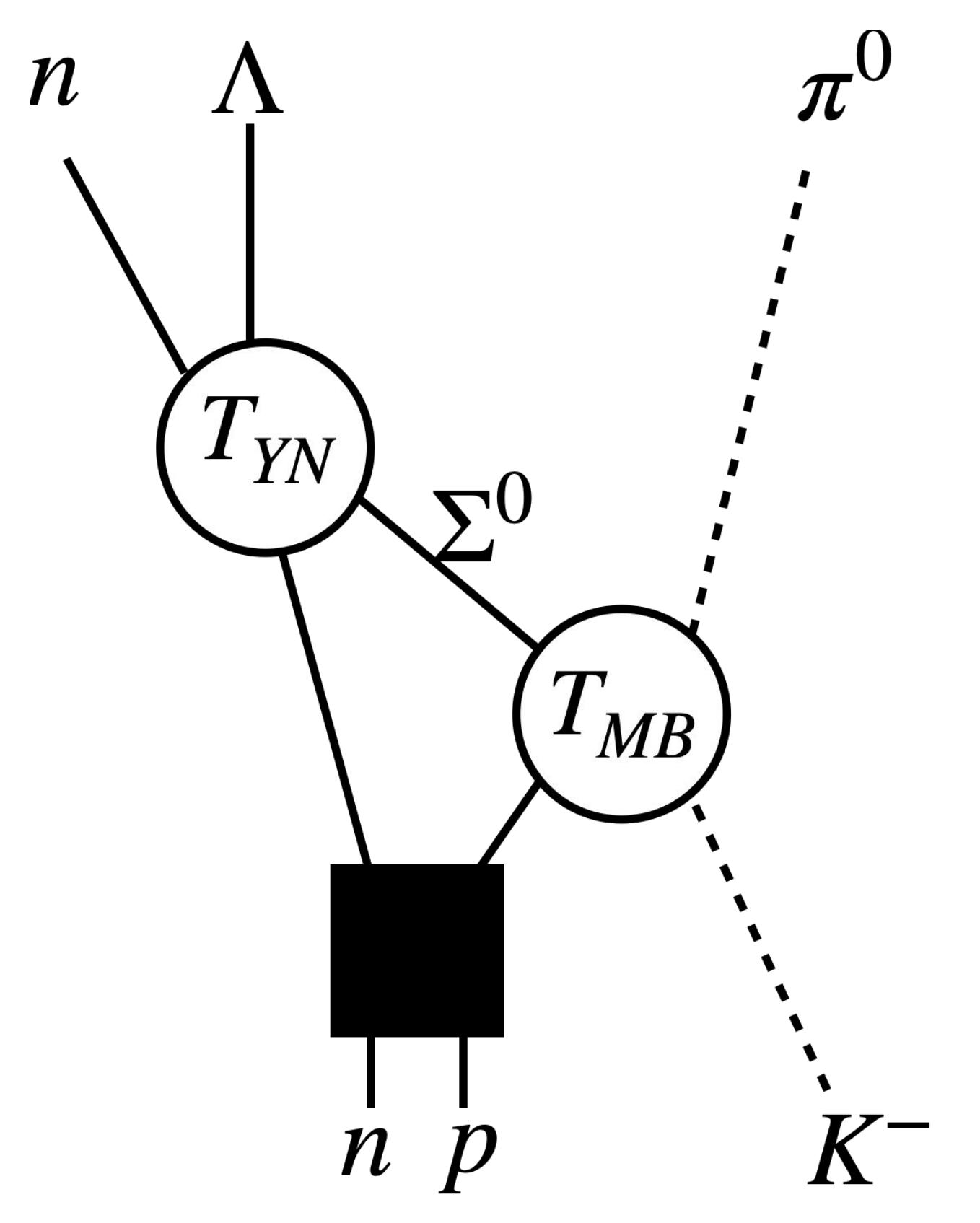}\\
        {\scriptsize Dia.~2a}
    \end{minipage}
    \begin{minipage}[b]{0.3\linewidth}
        \centering
        \includegraphics[width=2.5cm,clip]{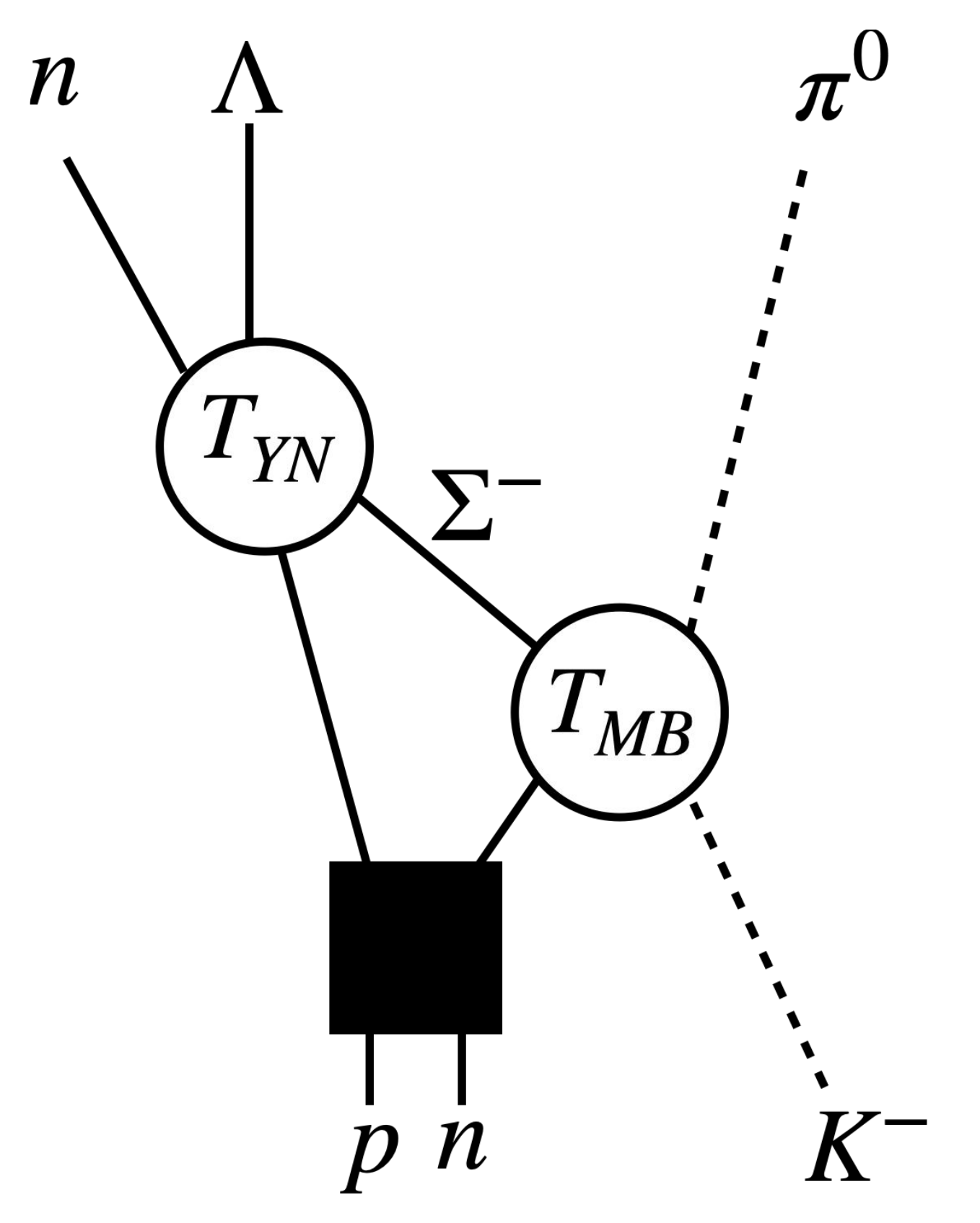}\\
        {\scriptsize Dia.~2b}
    \end{minipage}\\
    \vspace{0.2cm}
    \begin{minipage}[b]{0.3\linewidth}
        \centering
        \includegraphics[width=2.5cm,clip]{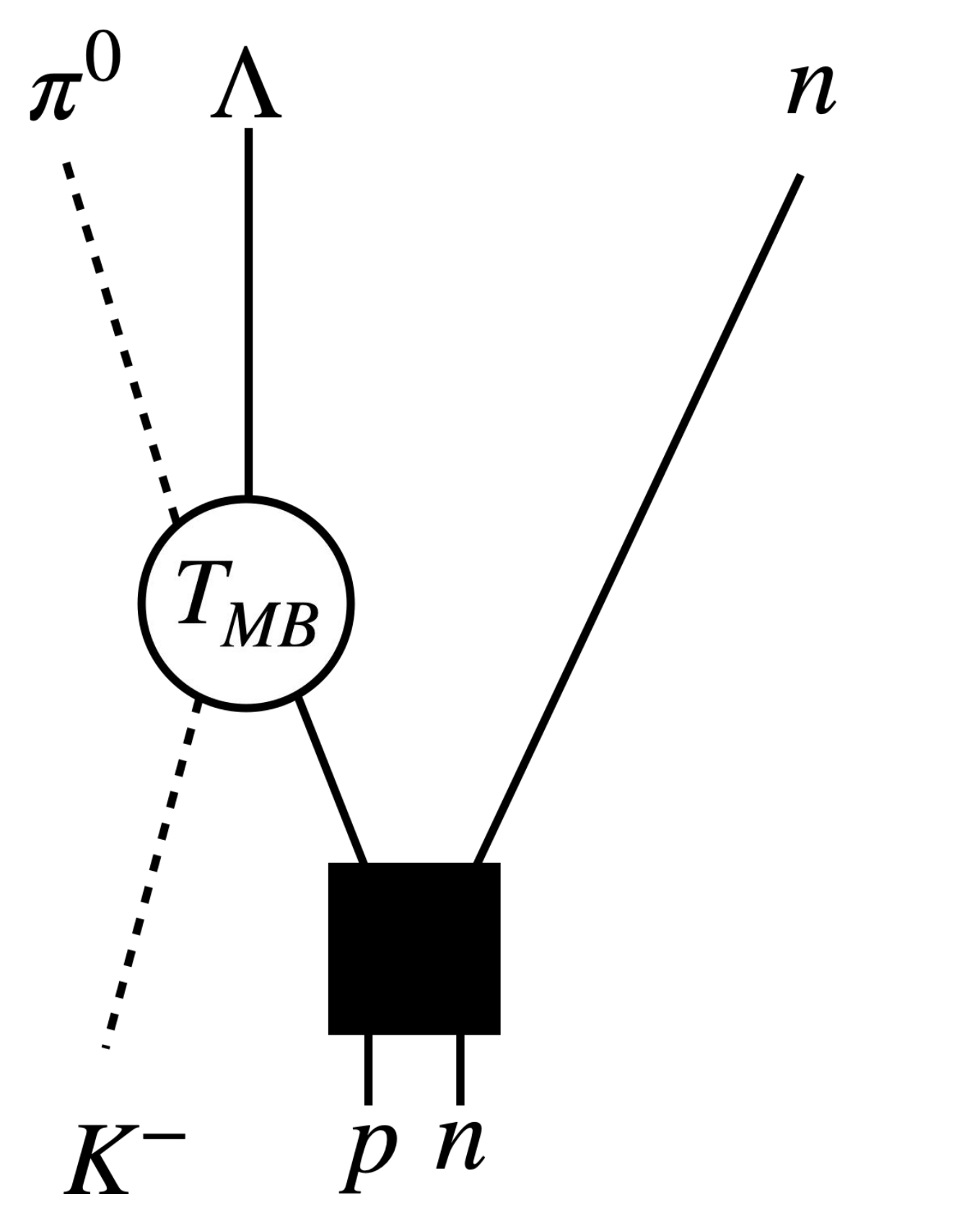}\\
        {\scriptsize Dia.~3}
    \end{minipage}
    \begin{minipage}[b]{0.3\linewidth}
        \centering
        \includegraphics[width=2.5cm,clip]{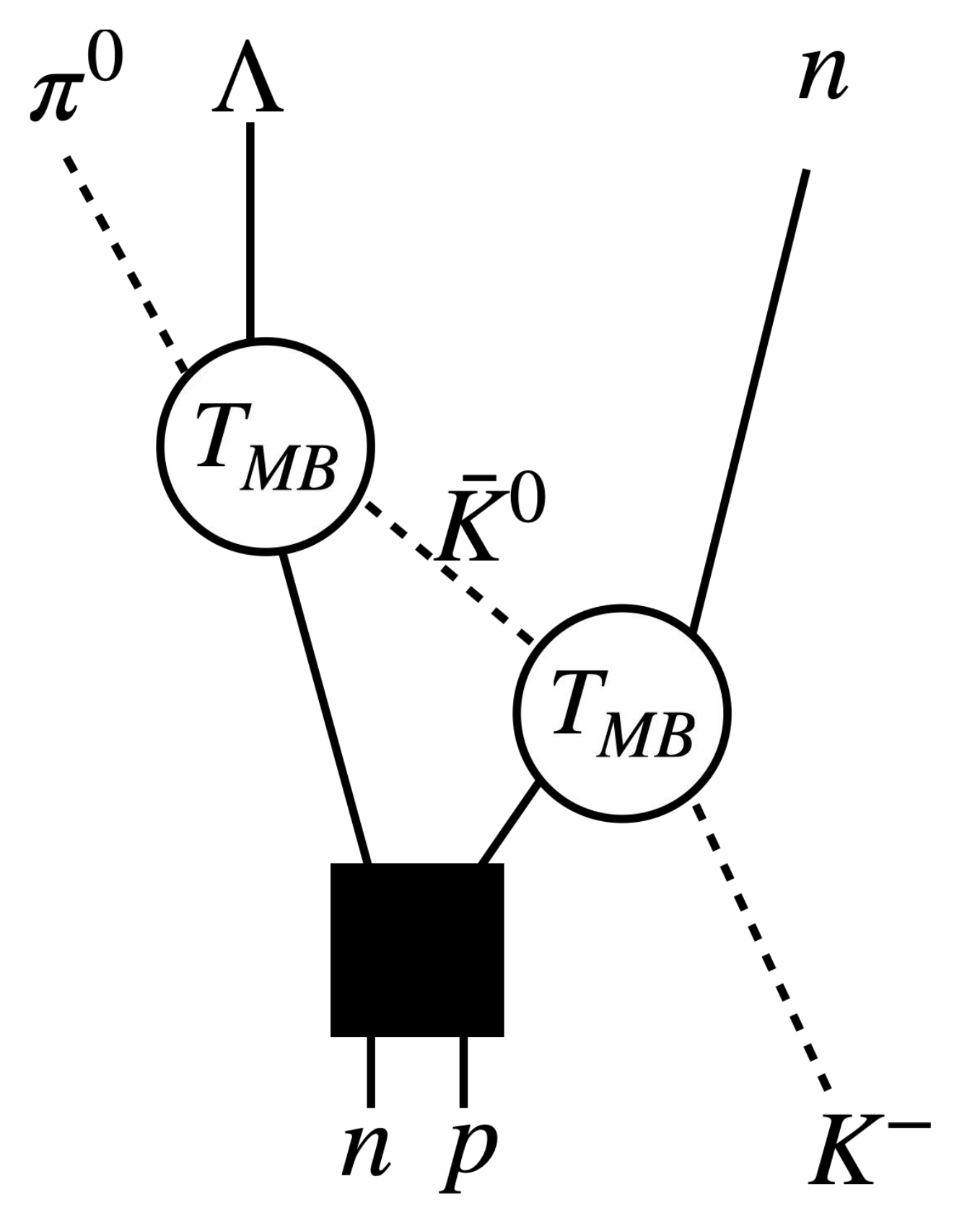}\\
        {\scriptsize Dia.~4a}
    \end{minipage}
    \begin{minipage}[b]{0.3\linewidth}
        \centering
        \includegraphics[width=2.5cm,clip]{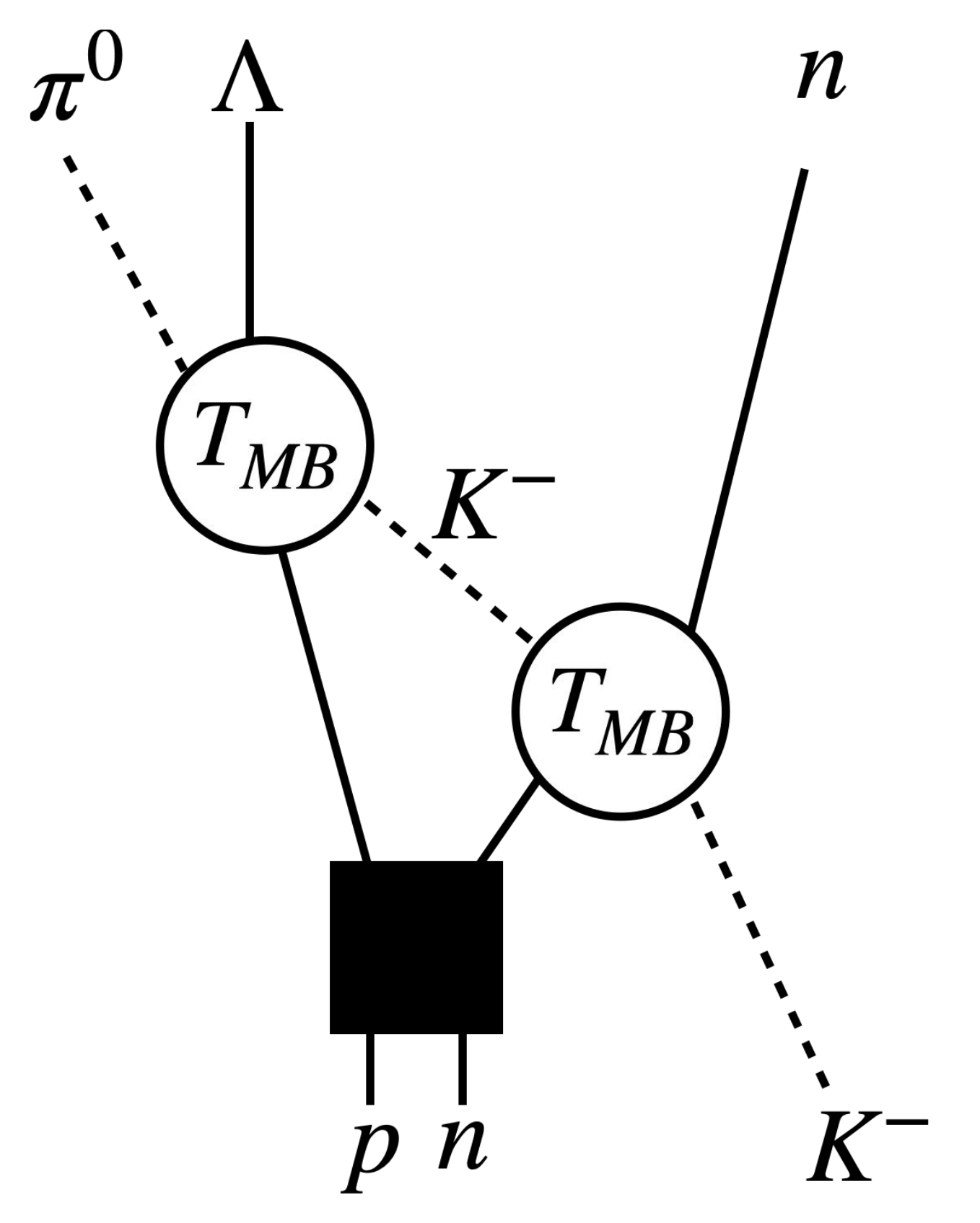}\\
        {\scriptsize Dia.~4b}
    \end{minipage}\\
    \vspace{0.2cm}
    \begin{minipage}[b]{0.3\linewidth}
        \centering
        \includegraphics[width=2.5cm,clip]{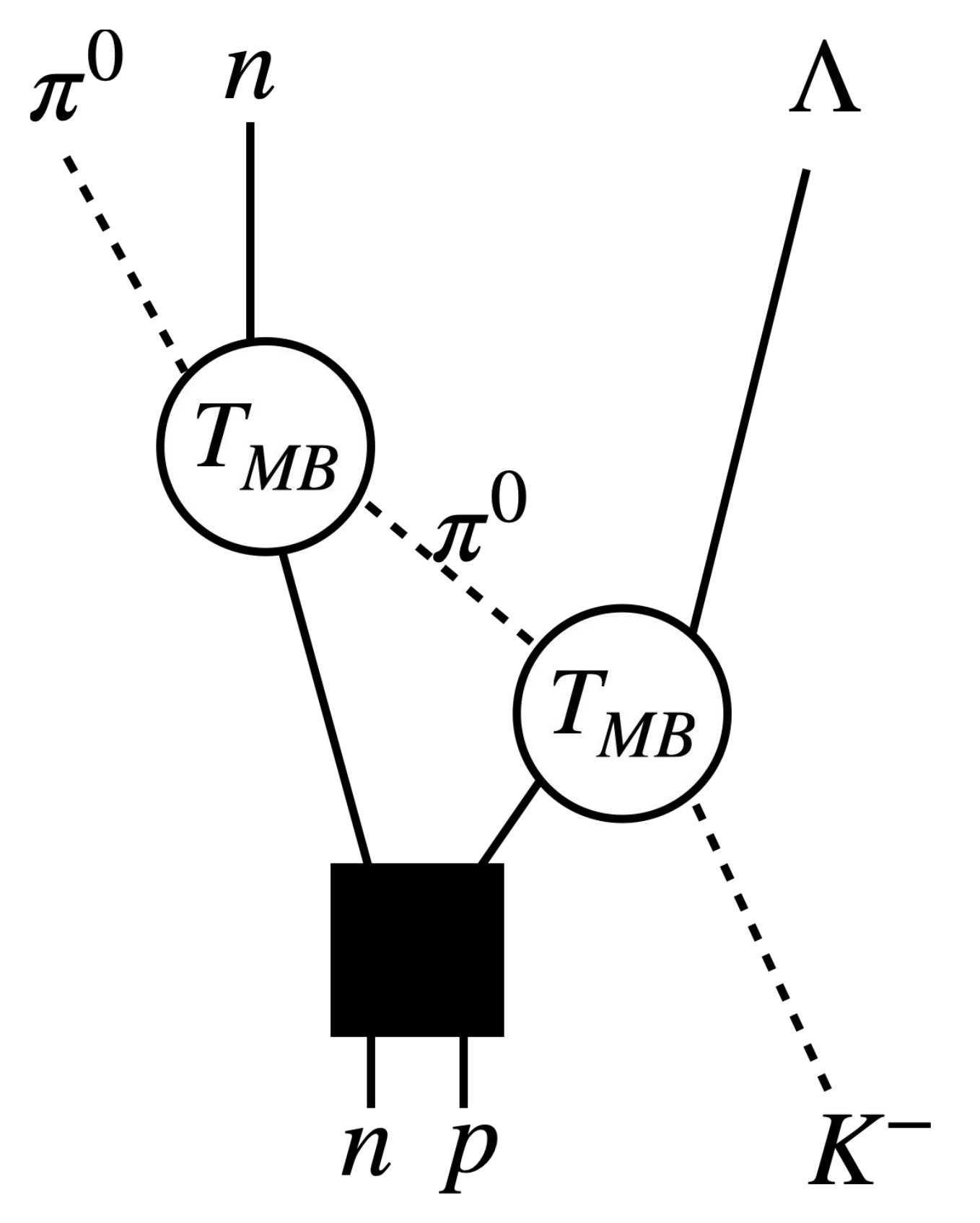}\\
        {\scriptsize Dia.~5a}
    \end{minipage}
    \begin{minipage}[b]{0.3\linewidth}
        \centering
        \includegraphics[width=2.5cm,clip]{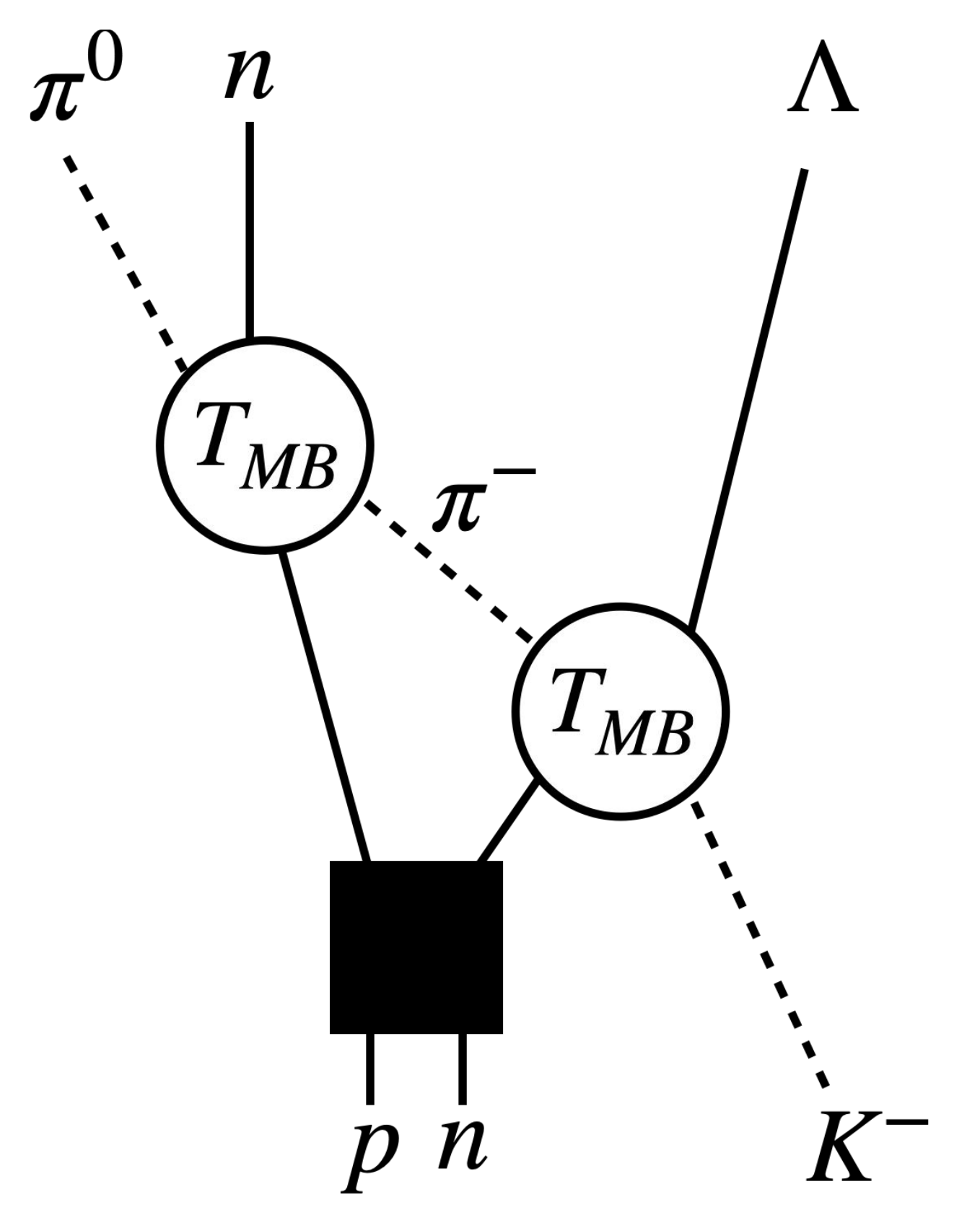}\\
        {\scriptsize Dia.~5b}
    \end{minipage}
    \caption{Same as in Fig.~\ref{fig:diangram_p} but for the $\Lambda n$ process.}
    \label{fig:diangram_n}
\end{figure}

\label{sec:3}
In this section, we formulate the scattering amplitudes of the $K^- d \rightarrow \pi \Lambda N$ reactions. The Feynman diagrams that we consider in our calculation are given in Fig.~\ref{fig:diangram_p} for the $\Lambda p$ process and Fig.~\ref{fig:diangram_n} for the $\Lambda n$ process. The calculation method is based on Refs. \cite{Jido:2010rx,Jido:2012cy,Yamagata-Sekihara:2012rij}. These diagrams contain absorption of the initial $K^-$ with one of the nucleons in the deuteron as $K^- N \to \pi Y$ or $K^- N \to \bar K N$, and the final state interaction between one of the particles emitted by the $K^-$ absorption and the other nucleon in the deuteron except for Dia.~3 (impulse approximation). The c.m.~energies of these processes are low enough and one may use $s$-wave amplitudes.
We use the hadron masses given in Ref.~\cite{ParticleDataGroup:2020ssz} in the following calculation.

First, we formulate the amplitude of the $\Lambda$ exchange diagrams Dia.~1 in Figs.~\ref{fig:diangram_p} and~\ref{fig:diangram_n}.
According to Ref.~\cite{Jido:2010rx}, we can write the scattering amplitude for the $\Lambda$ exchange diagram as

\begin{align}
    {\cal T}_{\pi \Lambda N}^{(1)} = & T_{\Lambda N}(M_{\Lambda N})
    \int \frac{d^{3}q}{(2\pi)^{3}} \frac{2M_\Lambda }{q^{2}-{M_\Lambda}^{2} + i\epsilon}
    \nonumber                                                       \\ &\times \tilde \varphi (\vec q+\vec p_{\pi})\ T_{K^{-}N \rightarrow \pi \Lambda}(W)
    \label{eq:dia1}
\end{align}
where $M_{\Lambda N}$ is the invariant mass of $\Lambda$ and $N$ in the final state, $q$ is the momentum of the exchange $\Lambda$, $\tilde \varphi$ is the $s$-wave deuteron wavefunction in the momenta space, and $W$ denotes the invariant mass of the initial kaon and the nucleon inside the deuteron. The amplitude $T_{\Lambda N}$ and $T_{K^- N \to \pi \Lambda}$ stand for the $s$-wave scattering amplitudes of the $\Lambda N \to \Lambda N$ and $K^- N \to \pi \Lambda$ processes, respectively. The charges of $N$ and $\pi$ are fixed in each process as shown in Figs.~\ref{fig:diangram_p} and~\ref{fig:diangram_n}. For instance, $K^- N \to \pi \Lambda$ for the $\Lambda p$ process corresponds to $K^- n \to \pi^- \Lambda$.

The nonrelativistic propagator is extended to a relativistic one as
\begin{align}
    \frac{1}{q_0 - \sqrt{\vec q^2+M^2}+i\epsilon} \simeq \frac{2M}{q^2-M^2 + i\epsilon}
    \nonumber
\end{align}
for simplicity of the calculation. With this approximation we can perform the momentum integral of Eq.~\eqref{eq:dia1} analytically, which reduces the calculation costs a lot.

For the baryon exchange diagrams, $q^0$ and $W$ are fixed as
\begin{align}
    q^0 & = M_{K^-} + M_d - \qty(M_2 - \frac{B_d}{2}) - p^0_\pi
    \label{eq:q0}
\end{align}
\begin{align}
    W & = \qty(M_1 - \frac{B_d}{2}) + M_{K^-}
    \label{eq:W}
\end{align}
where $B_d$ is the binding energy of deuteron, $M_1$($M_2$) is the participant (spectator) nucleon mass in $K^-N\to \pi\Lambda$, and $p^0_\pi$ is the energy of the final-state pion.

We neglect the small $d$-wave component of the deuteron wavefunction and use a parameterization of the $s$-wave component given by an analytic function in the CD-Bonn potential \cite{Machleidt:2000ge} as
\begin{align}
    \tilde \varphi(p) = N\sum^{11}_{j=1} \frac{C_j}{p^2+m^2_j}
    \label{deuteron}
\end{align}
where $N$ is the normalization factor, $C_j$ and $m_j$ were determined in Ref.~\cite{Machleidt:2000ge}. The argument of $\tilde \varphi$ in Eq.~\eqref{eq:dia1} is fixed by the momentum conservation of the vertex of the first-scattering.

Next, the $\Sigma$ exchange diagrams given by Dia.~2 in Figs.~\ref{fig:diangram_p} and~\ref{fig:diangram_n} are formulated as

\begin{align}
    {\cal T}_{\pi \Lambda N}^{(2)} = & \ T_{\Sigma N \rightarrow \Lambda N}(M_{\Lambda N})
    \int \frac{d^{3}q}{(2\pi)^{3}} \frac{2M_{\Sigma}}{q^{2}-{M_{\Sigma}}^{2} + i\epsilon}
    \nonumber                                                                                                            \\
                                     & \times \tilde \varphi (\vec q+\vec p_{\pi})\ T_{K^{-}N \rightarrow \pi \Sigma}(W)
\end{align}
as well as the foreground diagram. Here $q_0$ and $W$ are fixed as we do in Eqs. \eqref{eq:q0} and \eqref{eq:W}. The charge of the exchange $\Sigma$ is specified in Figs.~\ref{fig:diangram_p} and \ref{fig:diangram_n}.

Then we consider the amplitudes of the impulse approximation given by Dia.~3. The amplitudes are calculated as
\begin{align}
    {\cal T}_{\pi \Lambda N}^{(3)} = T_{K^{-}N \rightarrow \pi \Lambda}(M_{\pi\Lambda})\ \tilde \varphi (\vec p_{N})
\end{align}
where $\vec p_N$ is the momentum of the spectator nucleon in the rest frame of the deuteron.

The kaon exchange processes given by Dia.~4 can be calculated as
\begin{align}
    {\cal T}_{\pi \Lambda N}^{(4)} = & \ T_{\bar K N \rightarrow \pi \Lambda}(M_{\pi \Lambda})
    \int \frac{d^{3}q}{(2\pi)^{3}} \frac{\tilde \varphi (\vec q+\vec p_{N})}{q^{2}-{M^{2}_{\bar K}}+ i\epsilon}
    \nonumber                                                                                  \\ & \times
    T_{K^{-}N \rightarrow \bar K N}(W).
\end{align}
For $K^-d\rightarrow\pi^-\Lambda p$, the exchanged kaon is only $K^-$, however for $K^-d\rightarrow\pi^0\Lambda n$, $K^-$ and $\bar K^0$ are allowed as the exchanged kaon considering charge conversion.
For the kaon exchange diagrams, $q^0$ is fixed as
\begin{align}
    q^0  = M_{K^-} + M_d - \qty(M_2 - \frac{B_d}{2}) - p^0_N
\end{align}
with $p^0_N$ the energy of the final-state nucleon and $W$ is the same as Eq. \eqref{eq:W}.

Finally we obtain the amplitudes for the pion exchange processes given by Dia.~5 as
\begin{align}
    {\cal T}_{\pi \Lambda N}^{(5)} = & T_{\pi N \rightarrow \pi N}(M_{\pi N})
    \int \frac{d^{3}q}{(2\pi)^{3}} \frac{\tilde \varphi (\vec q+\vec p_{\Lambda})}{q^{2}-{M^{2}_{\pi}}+ i\epsilon}
    \nonumber                                                                 \\ & \times
    T_{K^{-}N \rightarrow \pi \Lambda}(W)
\end{align}
where both $\pi^0$ and $\pi^-$ are allowed as the exchanged pion for each process, $\vec p_\Lambda$ is the momentum of the final-state $\Lambda$ baryon in the total c.m. frame.
For the pion exchange diagrams, $q^0$ is fixed as
\begin{align}
    q^0  = M_{K^-} + M_d - \qty(M_2 - \frac{B_d}{2}) - p^0_\Lambda
\end{align}
with $p^0_\Lambda$ the energy of the final-state $\Lambda$ baryon and $W$ is the same as Eq. \eqref{eq:W}.

The total amplitude for $K^- d\rightarrow \pi \Lambda N$ is given by the sum of all amplitudes described above as
\begin{align}
    {\cal T}_{\pi \Lambda N} & = \sum_j {\cal T}_{\pi \Lambda N}^{(j)}.
\end{align}
The amplitude of primary interest (the foreground amplitude) in our calculation for $K^- d\rightarrow \pi \Lambda N$ is given by the $\Lambda$ exchange diagram (Dia.~1 in Figs.~\ref{fig:diangram_p} and \ref{fig:diangram_n}) as
\begin{align}
    {\cal T}^{\rm FG}_{\pi \Lambda N} & = {\cal T}_{\pi \Lambda N}^{(1)},
\end{align}
and we call the rest of the processes as background
\begin{align}
    {\cal T}^{\rm BG}_{\pi \Lambda N} & = \sum_{j\neq 1} {\cal T}_{\pi \Lambda N}^{(j)}.
\end{align}

The isospin breaking effects on the $T$-matrix of the reaction $\cal T$ are counted by using the observed masses for the exchanged particles and through the amplitudes of the absorption processes and the final state interactions. The details of isospin breaking of these amplitudes are described below.

\subsection{Hyperon-nucleon and meson-baryon amplitudes}
In this section, we explain the hyperon-nucleon and meson-baryon amplitudes that we use in the calculation of the cross section.
We parametrize the low-energy $s$-wave $\Lambda N$ scattering amplitude by the scattering length $a_{\Lambda N}$ and the effective range $r_{\Lambda N}$ given by
\begin{align}
    \label{eq:LN_amp}
    T_{\Lambda N}  =\  {\cal{N}}\frac{1}{-\displaystyle{\frac{1}{a_{\Lambda N}}}+\displaystyle{\frac{1}{2} r_{\Lambda N} {p^{\, *}_{\Lambda}}^2}-ip^{\, *}_{\Lambda}}
\end{align}
where $p^{\, *}_{\Lambda}$ is the momentum in the $\Lambda$-nucleon c.m. frame, and the kinematic factor $\cal N$ is given by
\begin{align}
    {\cal N}        =\ - \frac{8\pi M_{\Lambda N}}{\sqrt{(2M_Y)(2M_2)(2M_f)(2M_\Lambda)}}
    \label{eq:kf}
\end{align}
for the $Y N_2 \to \Lambda N_f $ process, where $M_Y$, $M_2$ and $M_f$ are the masses of $Y$, $N_2$ and $N_f$, respectively. This isospin symmetry breaking is also used for the $\Sigma N\to\Lambda N$ transition.

The experimentally obtained $a_{\Lambda N}$ and $r_{\Lambda N}$ of the spin-triplet $\Lambda$-proton are $a^t_{\Lambda p} = -1.56^{+0.19}_{-0.22}\fm$ and $r^t_{\Lambda p} = 3.7^{+0.6}_{-0.6}\fm$, respectively \cite{Budzanowski:2010ib}. On the other hand, the $\Lambda$-neutron interaction is not measured separately. The sensitivity of the $\Lambda N$ scattering parameters to the cross section of the reaction will be discussed in Sec.~\ref{results_C}.

For the $\Sigma N\rightarrow\Lambda N$ transition amplitude $T_{\Sigma N\to\Lambda N}$, we employ the unitarity of $S$-matrix in the isospin-doublet $\Lambda N$ and $\Sigma N$ channels. With the diagonal $\Lambda N$ and $\Sigma N$ amplitudes given, we utilize the unitarity of the $S$-matrix to determine the off-diagonal amplitude $\Sigma N\rightarrow\Lambda N$. The details of the calculations are given in Appendix~\ref{appendix_A}. We first adopt $a_{\Sigma N} = 1.68 - i 2.35\fm$ which is obtained by the Nijmegen NSC97f potential \cite{Rijken:1998yy}. We also compare the results with $a_{\Sigma N} = -3.83 - i 3.01$ fm taken from the \Julich~  potential \cite{Haidenbauer:2005zh}. In the following calculations, we introduce the known isospin breaking effect into the transition amplitude $T_{\Sigma N\to\Lambda N}$ through the kinematic factor $\cal N$ given in Eq.~\eqref{eq:kf}. Anyhow the isospin breaking effect on the transition amplitude are negligibly small in the cross sections around the $\Lambda N$ threshold.

\begin{figure}[htbp]
    \centering
    \includegraphics[width=8.6cm,clip]{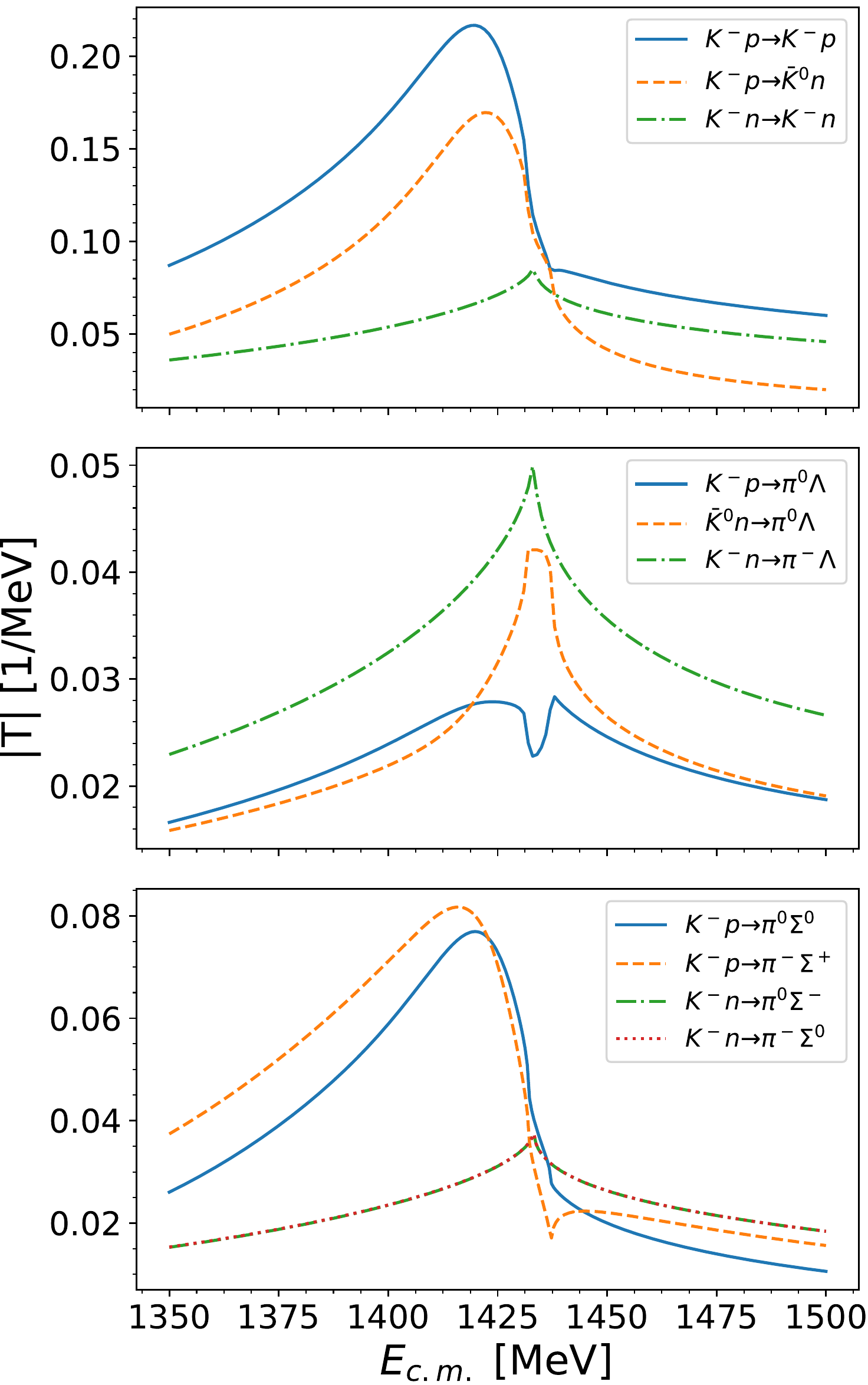}
    \caption{Modules of $\bar K N\rightarrow \bar K N$,  $\bar K N \rightarrow \pi \Lambda$ and $\bar K N \rightarrow \pi \Sigma$ scattering amplitude obtained in the chiral unitary approach using the parameters in Ref.~\cite{Oset:2001cn}}
    \label{fig:KN}
\end{figure}

The $\bar K N\rightarrow \bar K N$,  $\bar K N \rightarrow \pi \Lambda$ and $\bar K N \rightarrow \pi \Sigma$ amplitudes used are given by the chiral unitary model using the parameters in Ref.~\cite{Oset:2001cn} similarly to Refs.~\cite{Jido:2010rx,Jido:2012cy,Yamagata-Sekihara:2012rij}. The isospin breaking of these amplitudes is introduced by using the physical hadron masses in the loop functions and kinematic factors of the chiral unitary model. The interaction kernels are given by the Weinberg-Tomozawa interaction. They do not contain explicit flavor symmetry breaking. The subtraction constants are also determined in an isospin symmetric way. In Fig.~\ref{fig:KN}, we plot the modules of the $\bar K N$ scattering amplitudes used in this work. The isospin breaking of the $\bar KN$ thresholds are properly introduced by using the observed masses in the loop functions. Around the $\bar KN$ thresholds the isospin breaking effects look large. There, the theoretical description of the $\bar KN$ scattering amplitudes could be less reliable, because isospin breaking was not considered in the interaction kernels, although the isospin breaking effects may be enhanced around the thresholds.

For $\pi N\rightarrow \pi N$ scattering, we use the empirical amplitude $t_{\pi N}$\cite{piN} which is based on the available scattering data. It has been constructed isospin-symmetrically. Our amplitude $T_{\pi N}$ is obtained by $t_{\pi N}$ with a kinematic factor as follows:
\begin{align}
    T_{\pi N} = -\frac{8\pi M_{\pi N}}{\sqrt{k_i}\sqrt{k_f}\sqrt{2M_i}\sqrt{2M_f}} t_{\pi N}
\end{align}
where $M_{\pi N}$ in the invariant mass of $\pi N$, $k_i$($k_f$) is the momentum of the initial (final) pion, $M_i$($M_j$) is the mass of the initial (final) nucleon. The isospin breaking effect in the $\pi N$ amplitude $T_{\pi N}$ comes from the kinematic factor.

\section{Numerical results}
\label{results}

In this section, we show numerical results of the calculation for the $K^-d \rightarrow \pi^-\Lambda p$ and $K^-d \rightarrow \pi^0\Lambda n$ reactions. Using Eq. (\ref{cross section}), we evaluate the $\Lambda N$ invariant mass spectrum as
\begin{align}
    \label{cross section2}
    {\cal S}_N(M_{\Lambda N}) & \equiv k_{\rm c.m.}\frac{d\sigma}{d M_{\Lambda N}} \\ \nonumber
                              & =
    \frac{M_{d}M_{\Lambda} M_{N}}{16\pi^3E_{\rm c.m.}^{2}}
    \, |\vec p_{\pi}|\,
    |\vec p_{\Lambda}^{\, *}|\, \int |{\cal T}_{\pi \Lambda N}|^{2}
    d\cos\theta^*_{\Lambda}
\end{align}
with the $K^- d\rightarrow \pi \Lambda N$ scattering amplitude ${\cal T}_{\pi \Lambda N}$ discussed in the previous section. In our calculation, the incident kaon momentum in the laboratory frame is fixed at $0 \MeV/c$.

We use the most probable values of the observed spin-triplet $\Lambda$-proton scattering parameters $a_{\Lambda p} = -1.56\fm$ and $r_{\Lambda p} = 3.7\fm$ as the $\Lambda$-proton and $\Lambda$-neutron scattering amplitudes in Eq. \eqref{eq:LN_amp}, and use $a_{\Sigma N} = 1.68 - i 2.35$ (NSC97f).

\subsection{Background reduction}
\label{results_A}
\begin{figure}[htbp]
    \centering
    \includegraphics[width=8.6cm,clip]{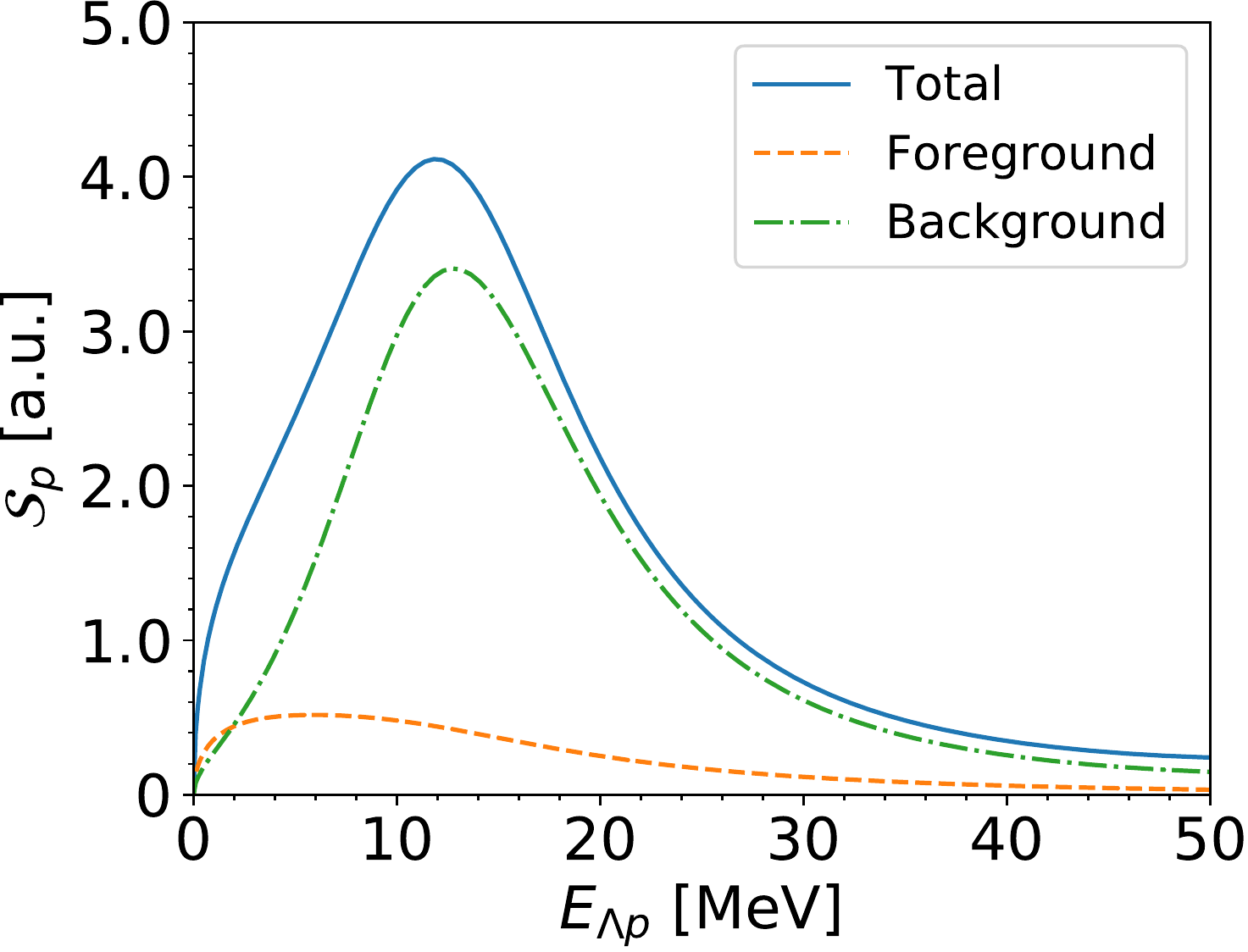}
    \caption{$\Lambda p$ invariant mass spectrum for the $\Lambda p$ process with stopped kaons. The horizontal axis represents the excitation energy from the threshold in a unit of$\MeV$. The solid, dashed, dash-dotted lines indicate the contributions of the total amplitudes, the foreground amplitude (only Dia.~1 in Fig.~\ref{fig:diangram_p}), the background amplitudes (the diagrams other than Dia.~1).
        % We use $a_{\Sigma N}= 1.68 - i 2.35\fm$ (NSC97f) for the $\Sigma N$ scattering length.
    }
    \label{fig:cross section_all}
\end{figure}

First we discuss background reduction in the $\Lambda N$ invariant mass spectrum for the $K^- d \to \pi \Lambda N$ reaction.
In Fig.~\ref{fig:cross section_all} we show the $\Lambda p$ invariant mass spectra for the $\Lambda p$ process where $\theta^*_\Lambda$ is integrated from $0$ to $\pi$.
Here the excitation energy $E_{\Lambda N}$ is defined by
\begin{align}
    E_{\Lambda N} \equiv M_{\Lambda N} - (M_\Lambda + M_N)
\end{align}
from the threshold is given instead of the mass itself.
We also plot the separated foreground and background spectra in Fig.~\ref{fig:cross section_all}. As seen in these plots the background contributions dominate the total spectra. Therefore it may be hard to extract the $\Lambda p$ scattering properties from the invariant spectra. We decompose them into components and look for appropriate kinematical conditions to reduce the background contributions.

\begin{figure}[htbp]
    \centering
    \includegraphics[width=8.6cm,clip]{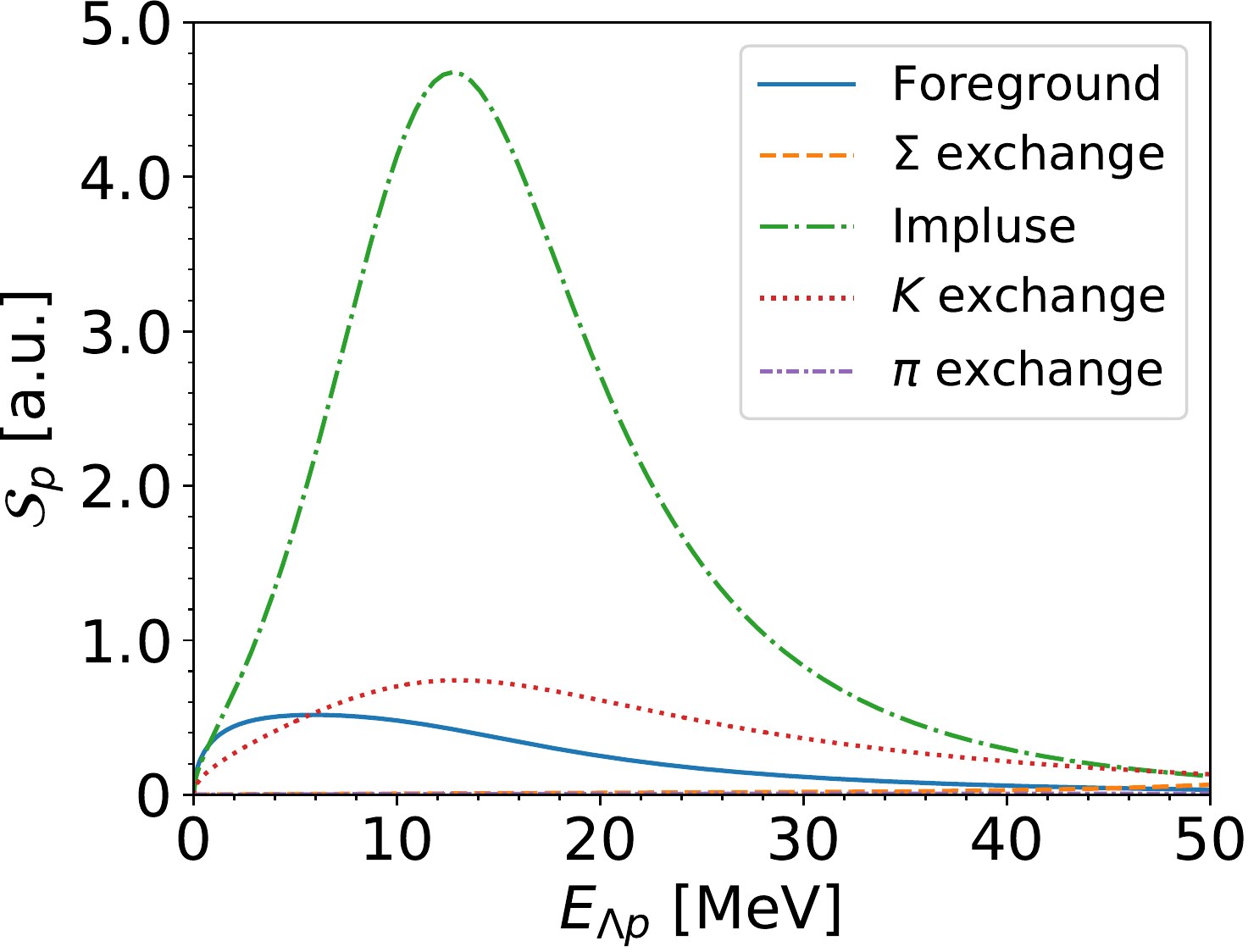}
    \caption{Decomposed background contributions for the $\Lambda p$ process.  The solid, dashed, dash-dotted, dotted, dash-dotted, density dash-dotted lines show the contributions from the foreground, $\Sigma$ exchange, impulse, $K$ exchange, $\pi$ exchange diagrams, respectively.
        % We use $a_{\Sigma N}= 1.68 - i 2.35\fm$ (NSC97f) for the $\Sigma N$ scattering length.
    }
    \label{fig:breakdown}
\end{figure}

In Fig.~\ref{fig:breakdown}, we show the decomposed components of the spectrum. As seen in the figure the impulse diagram gives the largest contribution and dominates the background. The contribution from the kaon exchange diagram is the second largest and is comparable with the foreground diagram. The $\Sigma$ and $\pi$ exchange diagrams give tiny contributions.
% For the excitation energy higher than $40\MeV$, the contribution from the $\Sigma$ exchange diagram becomes comparable to the foreground.

\begin{figure}
    \centering
    \includegraphics[width=8.6cm,clip]{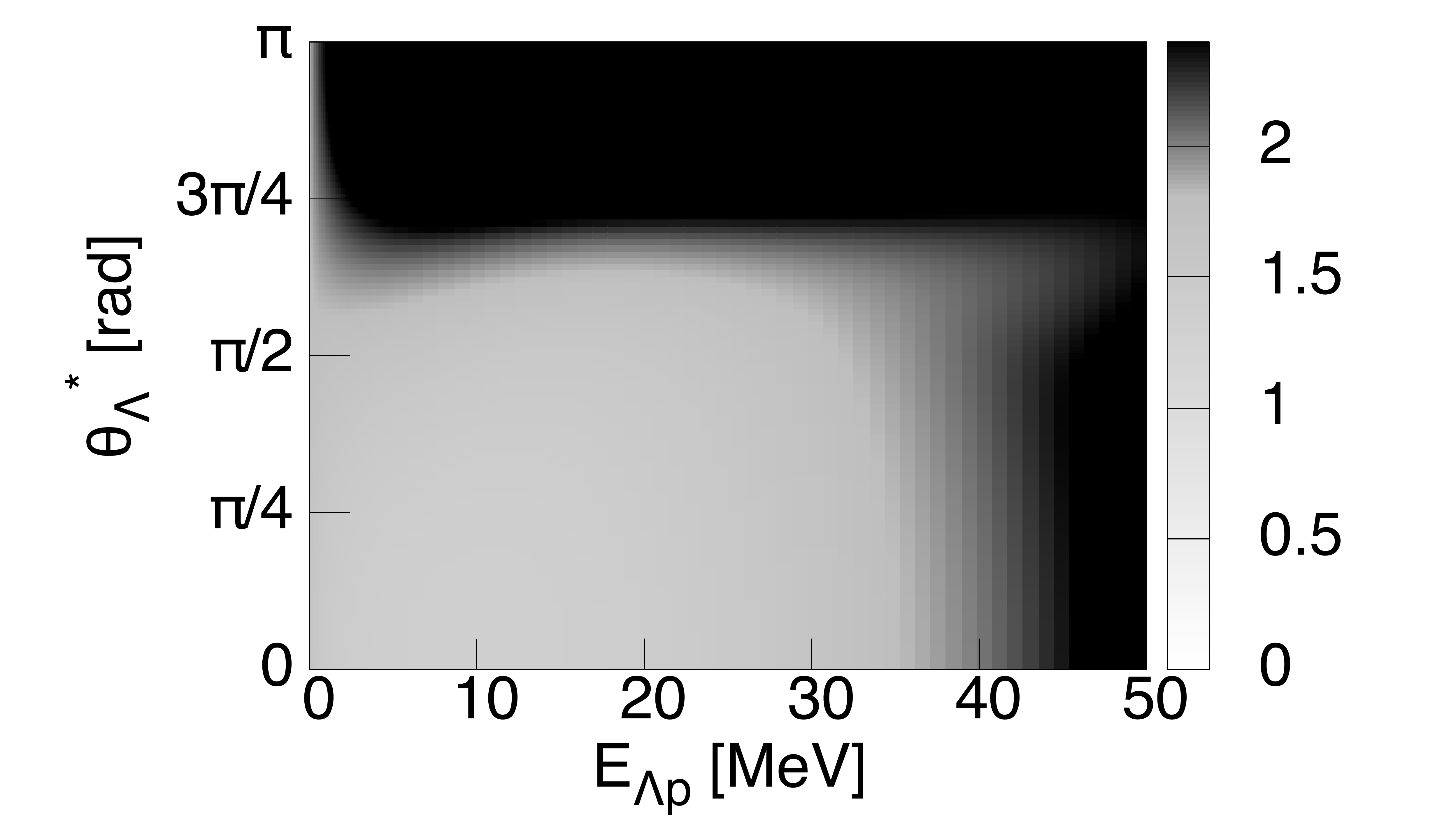}
    \caption{Two dimensional plot of $\mathcal{R}_B$ as a function
        of $\theta_\Lambda^*$ and $E_{\Lambda p}$ for the $\Lambda p$ process.
    }
    \label{fig:ratio_amp}
\end{figure}

Let us examine the angular dependence of the cross section for the impulse diagram. With a stopped kaon the final pion is emitted in the opposite direction to the $\Lambda$ in the laboratory frame, because the initial nucleon in the deuteron has a small Fermi momentum thanks to the small deuteron binding energy.
% The angle of the emitted $\Lambda$ in the c.m. frame of the $\Lambda N$ system, $\theta_\Lambda^*$, is defined by taking the pion momentum as a reference. 
Thus, the impulse diagram gives a larger contribution for larger $\theta_\Lambda^*$. The kaon exchange diagram has also similar angular dependence because the exchange kaon also has a small momentum for a stopped initial kaon. Therefore, the main background diagrams, Dias.~3 and 4, have a smaller contribution for smaller $\theta_\Lambda^*$. This can be checked by plotting a ratio defined by
\begin{align}
    \mathcal{R}_B = \frac{|{\cal T}^{\rm BG}_{\pi \Lambda N}|^2}{|{\cal T}^{\rm FG}_{\pi \Lambda N}|^2}.
    \label{eq:ratio_amp}
\end{align}
As seen in Fig.~\ref{fig:ratio_amp}, $\mathcal{R}_B$ is large at $\theta_\Lambda^* > 3\pi/4$ except near the threshold. In order to reduce the background, we should avoid this region.
In $E_{\Lambda N} \gtrsim 40 \MeV$, $\mathcal{R}_B$ is large independently of $\theta^*_{\Lambda}$. This is because the contribution of $\Sigma$ exchange increases as $E_{\Lambda N}$ approaches the $\Sigma N$ threshold seen in Fig.~\ref{fig:breakdown}.

\begin{figure}[htbp]
    \centering
    \includegraphics[width=8.6cm,clip]{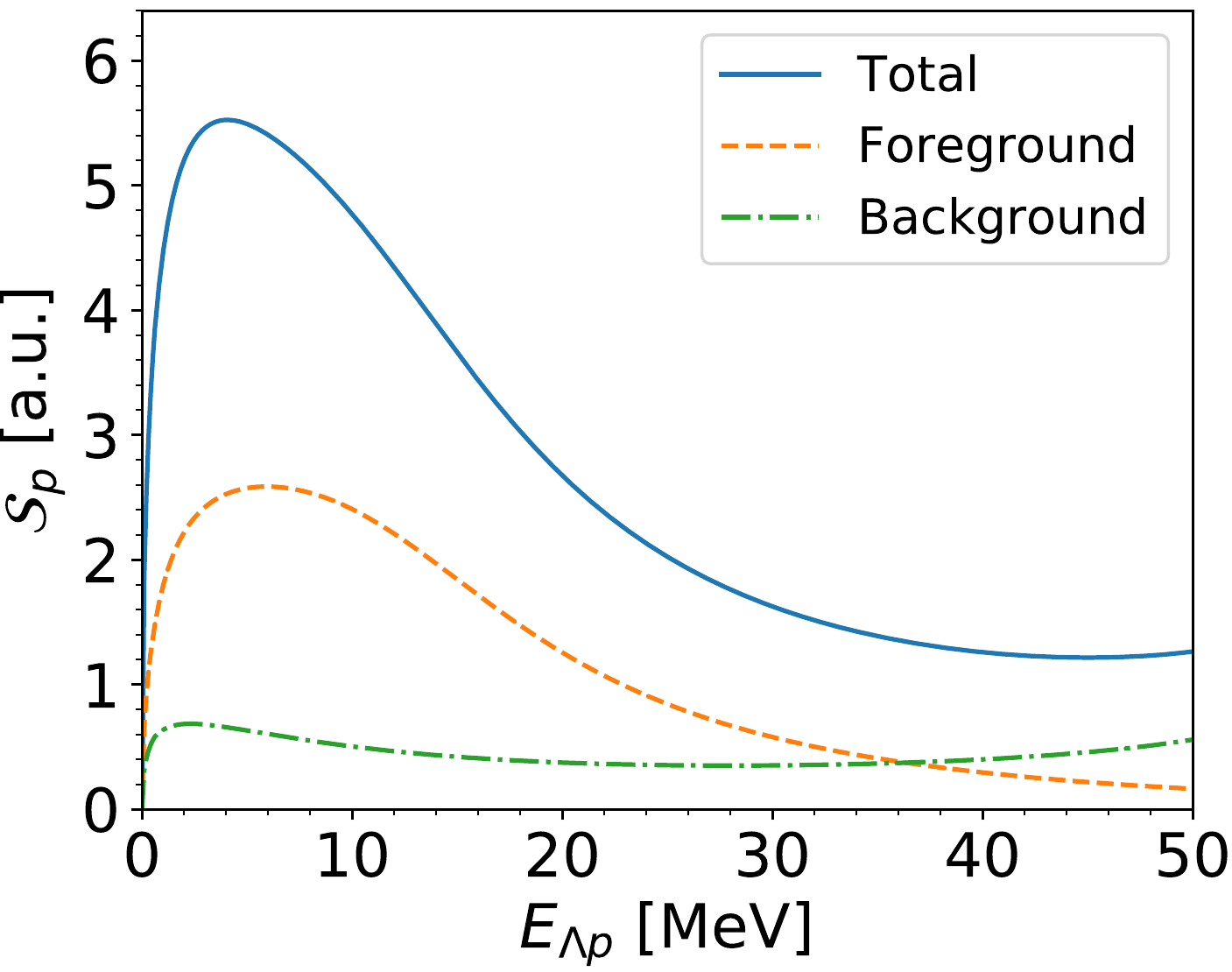}
    \caption{$\Lambda p$ mass spectra for the $\Lambda p$ process integrated in the range $[0,\pi/2]$ with stopped kaon. The horizontal axis represents the excitation energy from the threshold in a unit of$\MeV$. The solid, dashed, dash-dotted lines indicate the contributions of the total amplitudes, the foreground amplitude (only Dia.~1 in Fig.~(\ref{fig:diangram_p})), the background amplitudes (the diagrams other than Dia.~1).
    }
    \label{fig:cross section_limited}
\end{figure}

\begin{figure}[htbp]
    \centering
    \includegraphics[width=8.6cm,clip]{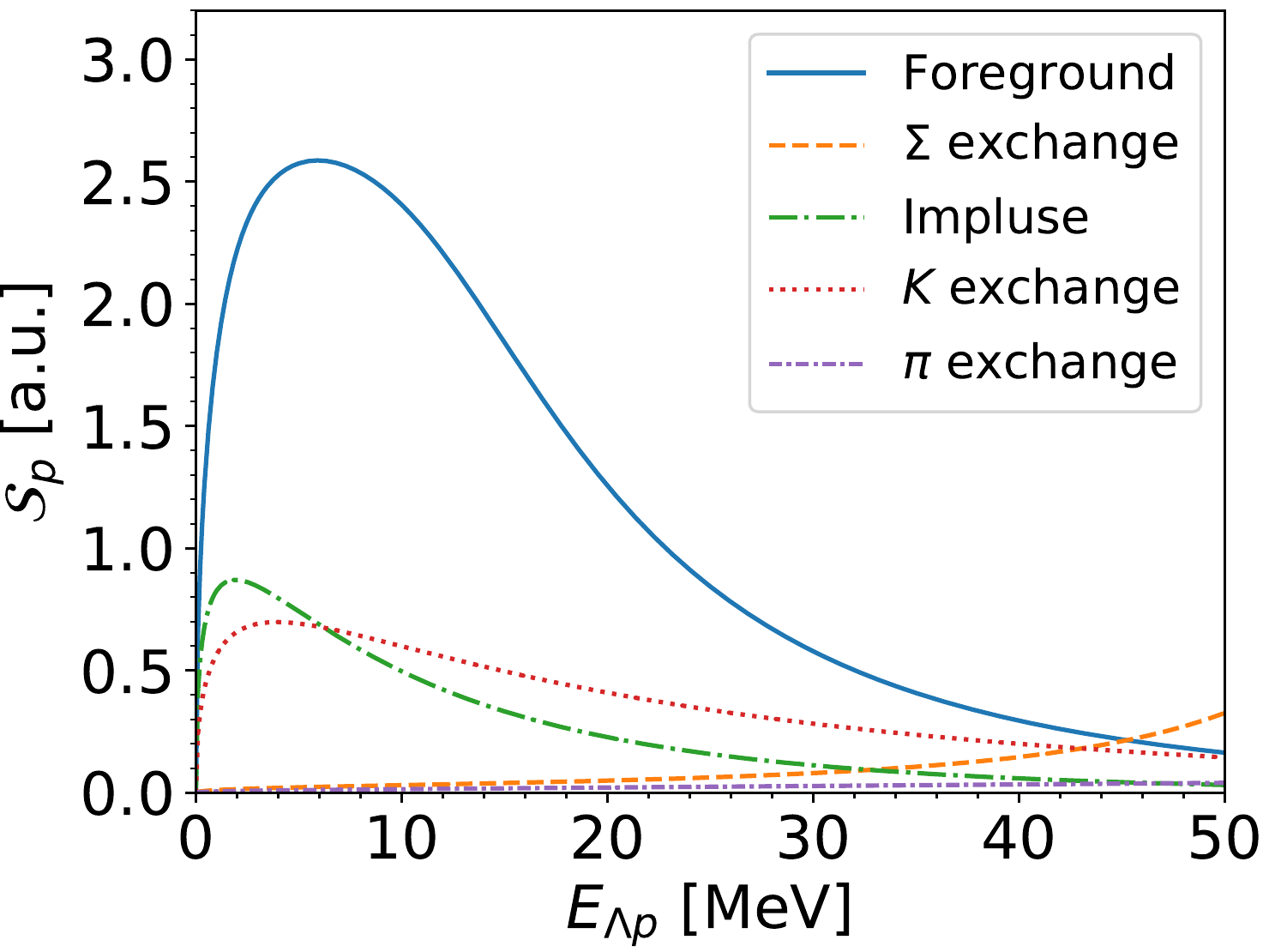}
    \caption{Decomposed background contributions for the $\Lambda p$ process integrated in the range $[0,\pi/2]$. The solid, dashed, dash-dotted, dotted, dash-dotted, density dash-dotted lines show the contributions from the foreground, $\Sigma$ exchange, impulse, $K$ exchange, $\pi$ exchange diagrams, respectively.
    }
    \label{fig:breakdown_2}
\end{figure}

In order to reduce the background contributions, let us propose to restrict the integral region of the angle $\theta^*_{\Lambda}$ as $0$ to $\pi/2$. In Fig.~\ref{fig:cross section_limited}, we show the total, foreground, and background $\Lambda N$ mass spectra for the $\Lambda p$ process integrated in the range $[0, \pi/2]$. From Fig.~\ref{fig:cross section_limited}, it can be seen that the background contributions are substantially suppressed for $E_{\Lambda p} < 30\MeV$. We also plot the the $\Lambda N$ mass spectrum for each contribution in Fig.~\ref{fig:breakdown_2}, showing that the foreground contribution dominates over the other contributions for $E_{\Lambda p} < 40\MeV$. Especially the contribution of the pion exchange diagram is negligibly small. The $\Sigma$ exchange contribution is also quite small $E_{\Lambda p} < 30\MeV$, but it gets comparable to the foreground contribution for $E_{\Lambda p} > 40\MeV$. Hereafter the integral of $\theta^*_{\Lambda}$ is performed from $0$ up to $\pi/2$.

The purpose of this study is to see isospin symmetry breaking in the $\Lambda N$ interaction. It is very important to control the isospin symmetry breaking effects from the other sources. In particular, there is a large isospin breaking effect around the $\bar KN$ thresholds in the $\bar K N \to M B$ amplitudes as seen in Fig.~\ref{fig:KN}. If possible, it is better to avoid these energy region by controlling the kinematical variables of the final state.
In the diagrams expect the impulse approximation (Dia.~3) in Figs.~\ref{fig:diangram_p} and~\ref{fig:diangram_n}, the energy of the first scattering is determined by those of the initial kaon and the participant nucleon in the deuteron. Thus, the first scattering cannot be controlled by the final state kinematics. For the second scattering, on the other hand, the c.m. energy is dependent on the kinematics of the final state and is controllable.

\begin{figure}[htbp]
    \centering
    \includegraphics[width=8.6cm,clip]{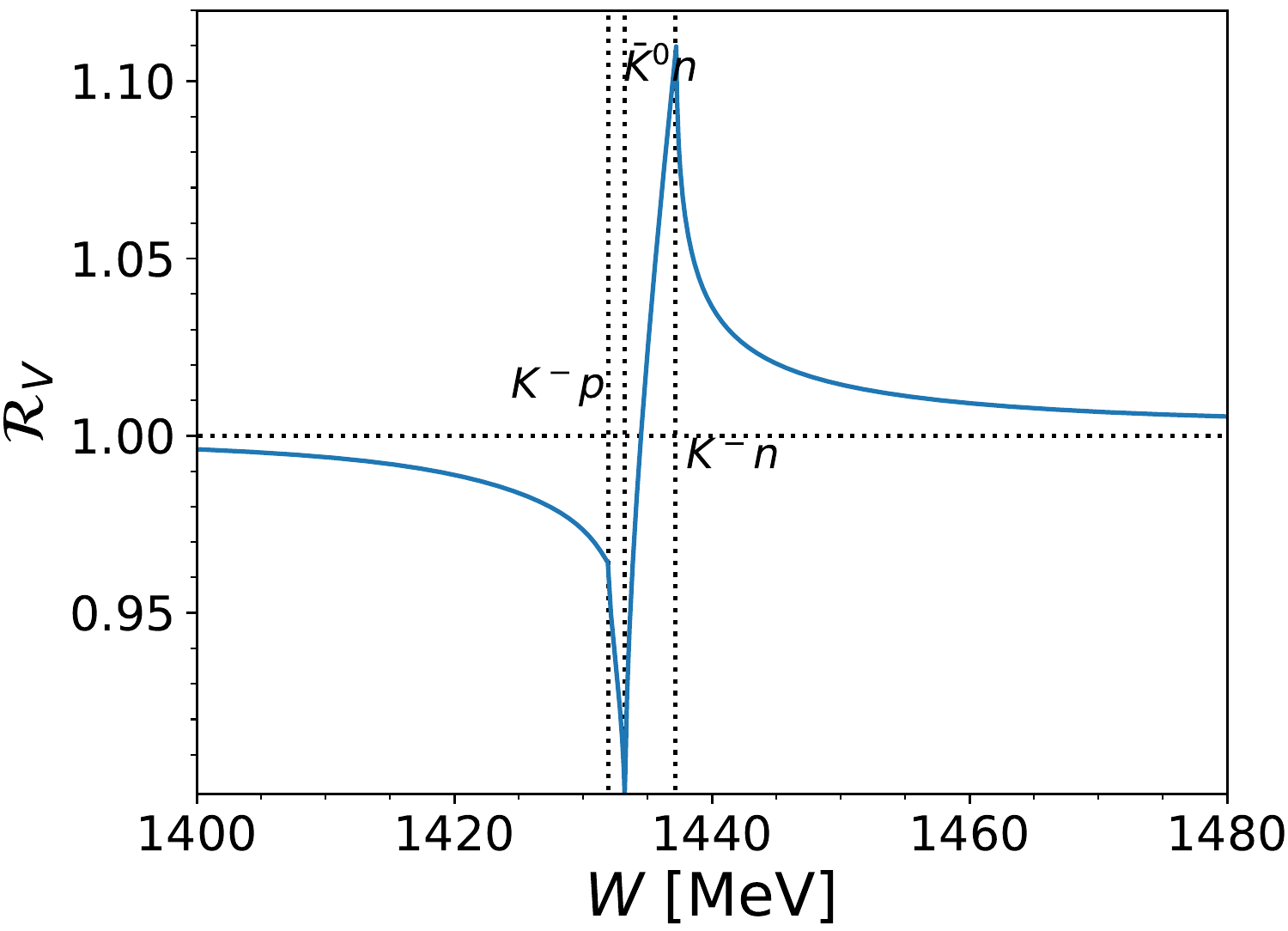}
    \caption{Ratio \eqref{eq: ratio_of_KNtopiL} for the $\bar K N \to \pi \Lambda$ amplitudes as a function of the c.m. energy of $\pi\Lambda$. The thresholds of $K^- p$, $\bar K^0 n$ and $K^- n$ are at $1431.95$, $1433.24$ and $1437.18$ MeV, respectively.}
    \label{fig:chum_ratio}
\end{figure}

\begin{figure}[htbp]
    \centering
    \includegraphics[width=8.6cm,clip]{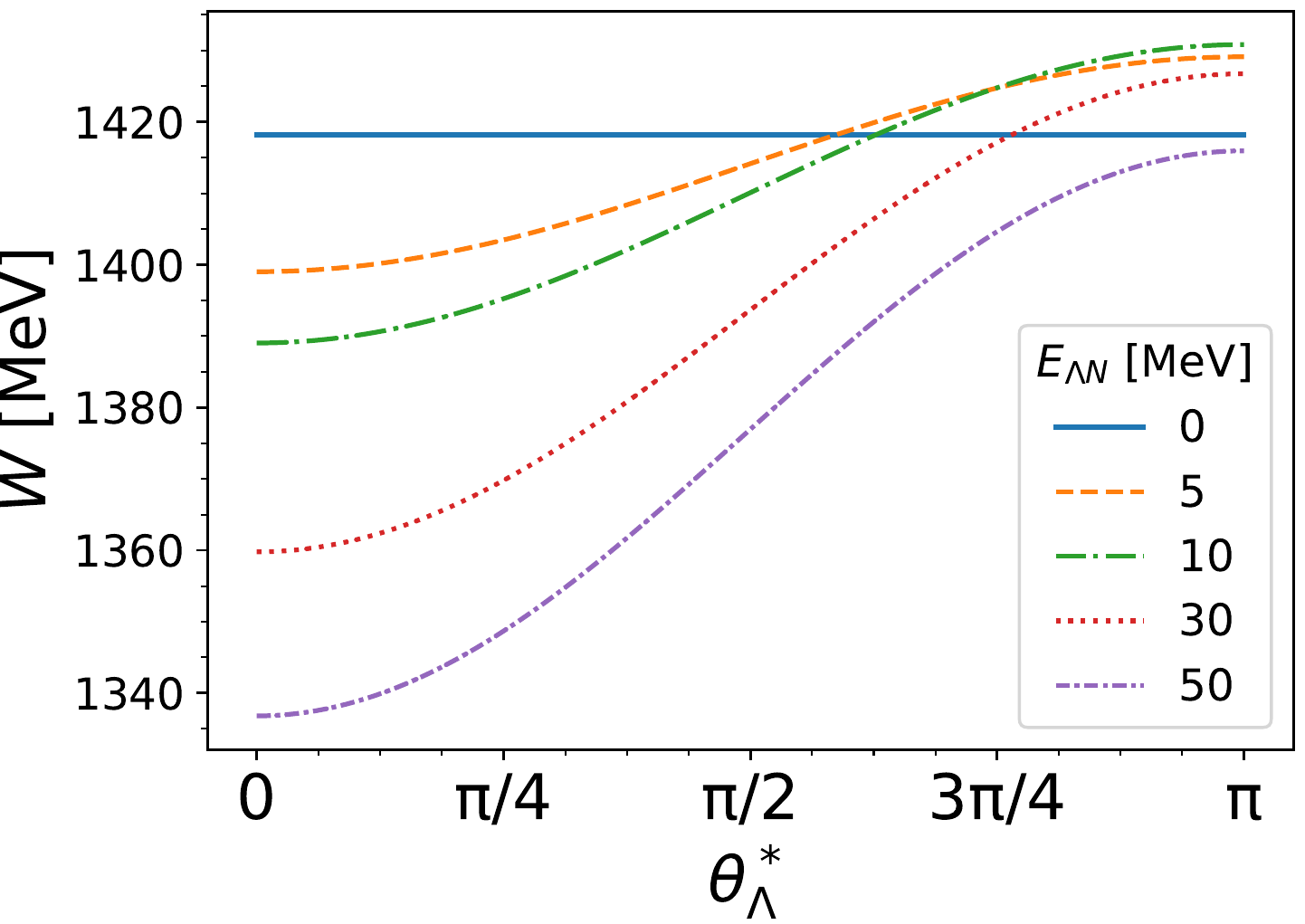}
    \caption{Center-of-mass energy $W$ of $\pi \Lambda$ as a function of $\theta_\Lambda^*$ for several $E_{\Lambda p}$. }
    \label{fig:MpiL_cos}
\end{figure}

In order to see isospin breaking of the $\bar K N \to \pi \Lambda$ amplitudes,
let us plot the following ratio of the $\bar K N \to \pi \Lambda$ amplitudes in Fig.~\ref{fig:chum_ratio}:
\begin{align}
    \mathcal{R}_V(W) = \frac{|T_{K^- p\rightarrow \pi^0\Lambda} - T_{\bar K^0 n\rightarrow \pi^0\Lambda}|}{\sqrt{2}|T_{K^- n\rightarrow \pi^-\Lambda}|}.
    \label{eq: ratio_of_KNtopiL}
\end{align}
The ratio should be unity if the amplitude is isospin symmetric. This figure shows the large isospin breaking effect around the threshold region, $1420 \MeV < W < 1450 \MeV$. The c.m. energy of the second scattering is determined by the final state momenta.
In order to find kinematic conditions in the final state corresponding a large isospin breaking effect in the $\bar K N \to \pi \Lambda$ amplitudes, we show the c.m. energy $W$ of $\pi \Lambda$ as  function of $\theta^*_\Lambda$ in Fig.~\ref{fig:MpiL_cos} for several $E_{\Lambda p}$.
One can see that if one wants to avoid the large isospin breaking area $1420 \MeV < W < 1450 \MeV$ for the $\bar K N \to \pi \Lambda$ amplitudes, smaller angles are favorable, such as $\theta^*_\Lambda < \pi/2$. This implies that our integral region of $\theta_\Lambda^*$ is in this safe range.

\subsection{Sensitivity to $a_{\Lambda N}$, $r_{\Lambda N}$ and $a_{\Sigma N}$}
\label{results_B}
In this section, we discuss the sensitivity of $a_{\Lambda N}$, $r_{\Lambda N}$ and $a_{\Sigma N}$.

\begin{figure}[htbp]
    \includegraphics[width=8.6cm,clip]{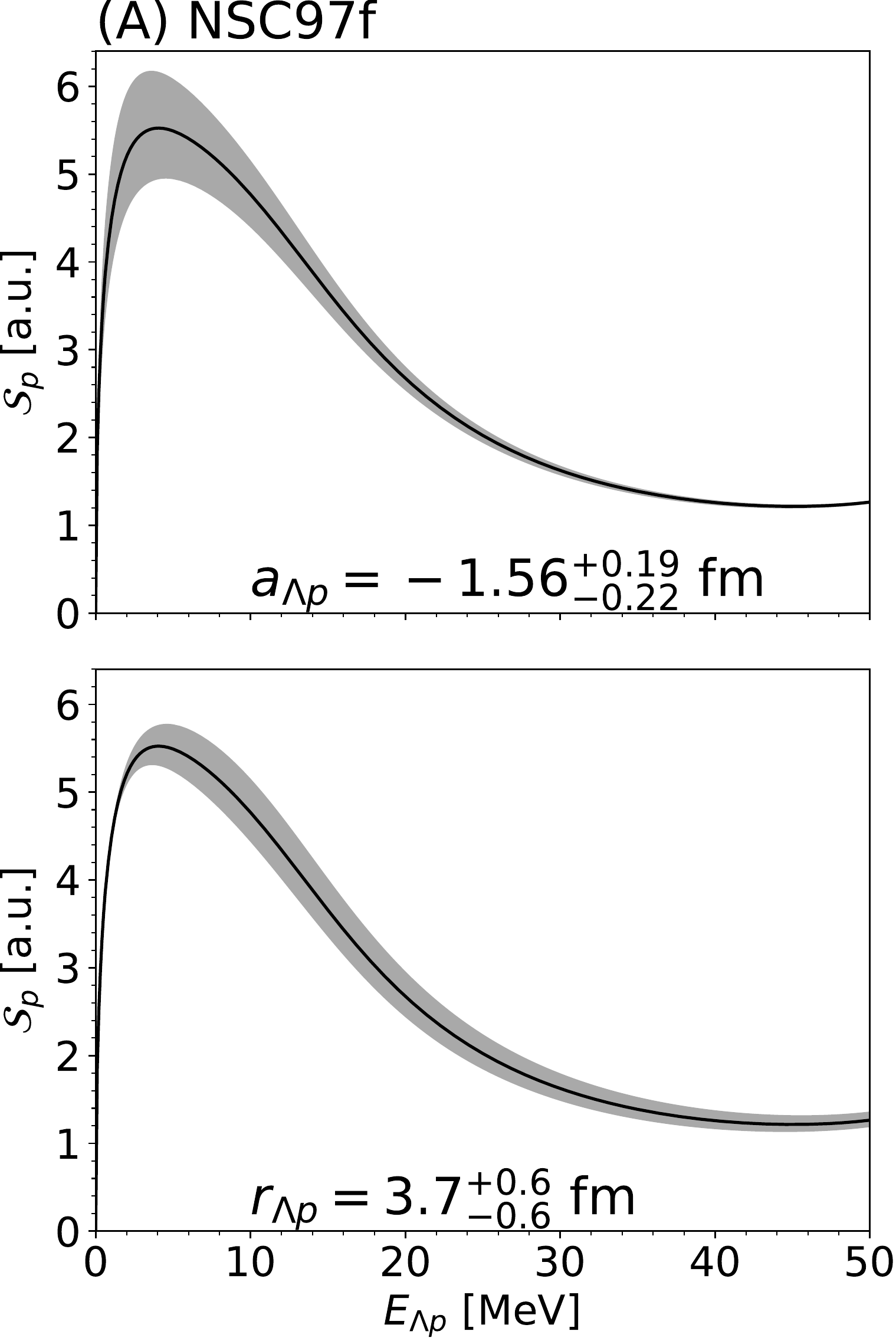}
    \caption{$\Lambda p$ invariant mass spectra calculated with different $a_{\Lambda p}$ and $r_{\Lambda p}$ values for the $\Lambda p$ process. In the upper plot the value of the scattering length is changed within $a_{\Lambda p}=-1.56^{+0.19}_{-0.22}\fm$, while the value of the effective range varies within $r_{\Lambda p}=3.7^{+0.6}_{-0.6}\fm$ in the lower plot.
    For the two plots $a_{\Sigma N}= 1.68 - i 2.35\fm$ (NSC97f) is used.
    }
    \label{fig:sensitivity_p_A}
\end{figure}

\begin{figure}[htbp]
    \includegraphics[width=8.6cm,clip]{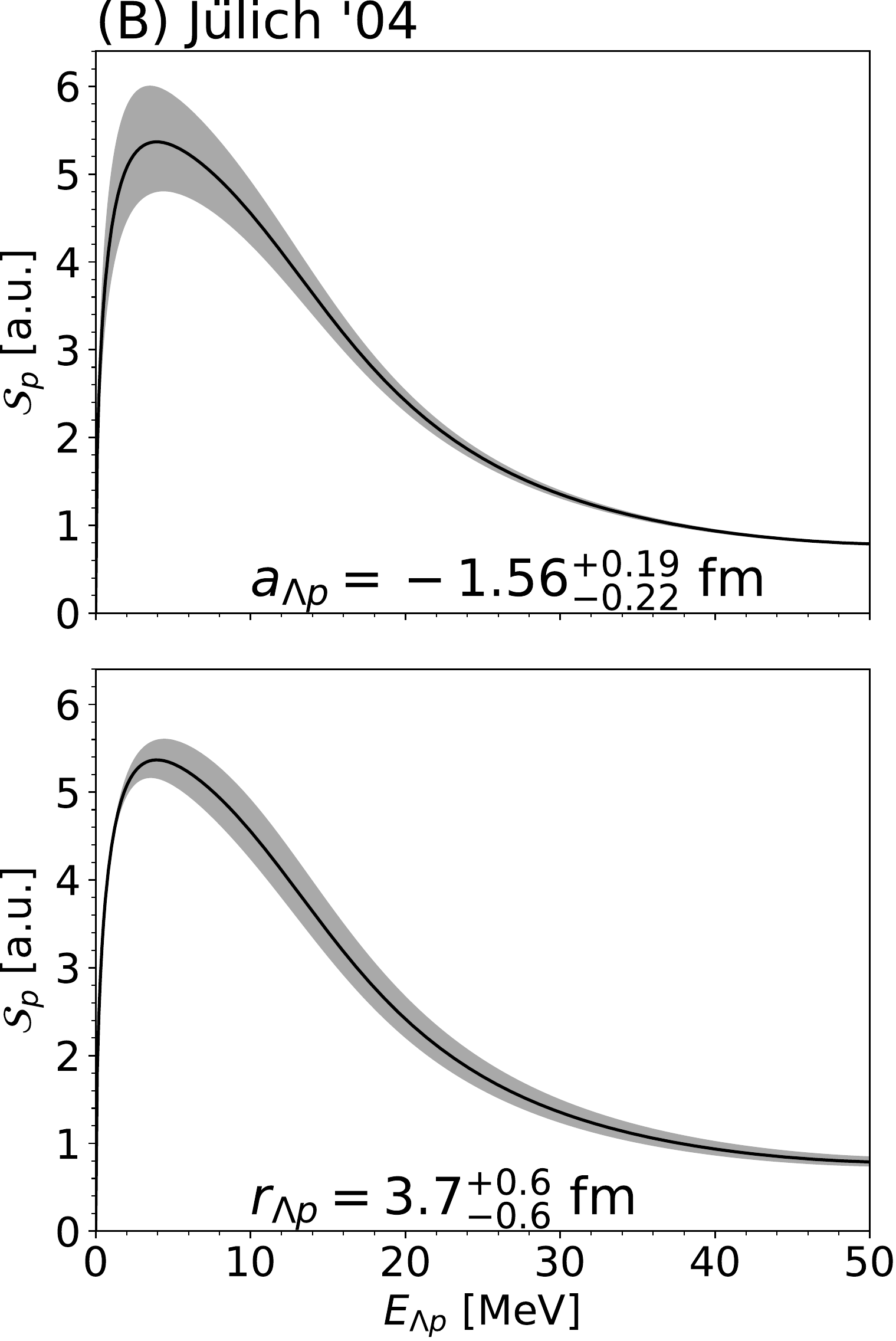}
    \caption{Same as Fig.~\ref{fig:sensitivity_p_A} but $a_{\Sigma N} = - 3.83 - i 3.01\fm$ (\Julich) is used.
    }
    \label{fig:sensitivity_p_B}
\end{figure}

First we evaluate the $\Lambda p$ invariant mass spectra by changing the values of $a_{\Lambda p}$ and $r_{\Lambda p}$ within the experimental errors in order to see the experimental uncertainties on the $\Lambda p$ scattering parameters. The result is shown in Fig.~\ref{fig:sensitivity_p_A} using $a_{\Sigma N}= 1.68 - i 2.35 \fm$ (NSC97f) and Fig.~\ref{fig:sensitivity_p_B} using $a_{\Sigma N} = - 3.83 - i 3.01\fm$ (\Julich).
In each upper plot we change the value of the scattering length within $a_{\Lambda p}=-1.56^{+0.19}_{-0.22}\fm$ with fixing the effective range as $r_{\Lambda p} = 3.7 \fm$, while we vary the value of the effective range within $r_{\Lambda p}=3.7^{+0.6}_{-0.6}\fm$ with $a_{\Lambda p} = -1.56 \fm$ in each lower plot.
From these plots one can see that the $\Lambda p$ mass spectrum changes in the regions of the lower excitation energies $0 < E_{\Lambda p} < 15\MeV$ when $a_{\Lambda p}$ is varied, and it changes in $5 < E_{\Lambda p} < 30\MeV$ when $r_{\Lambda p}$ is varied.

Comparing Figs.~\ref{fig:sensitivity_p_A} and \ref{fig:sensitivity_p_B}, we find that at low energies the two models have little difference, while at $E_{\Lambda p} > 40\MeV$ the spectrum using $a_{\Sigma N}= 1.68 - i 2.35 \fm$ (NSC97f) increases as increase of $E_{\Lambda p}$ and that using $a_{\Sigma N} = - 3.83 - i 3.01\fm$ (\Julich) does not.

\begin{figure}[htbp]
    % \begin{minipage}[b]{0.49\linewidth}
    \includegraphics[width=8.6cm,clip]{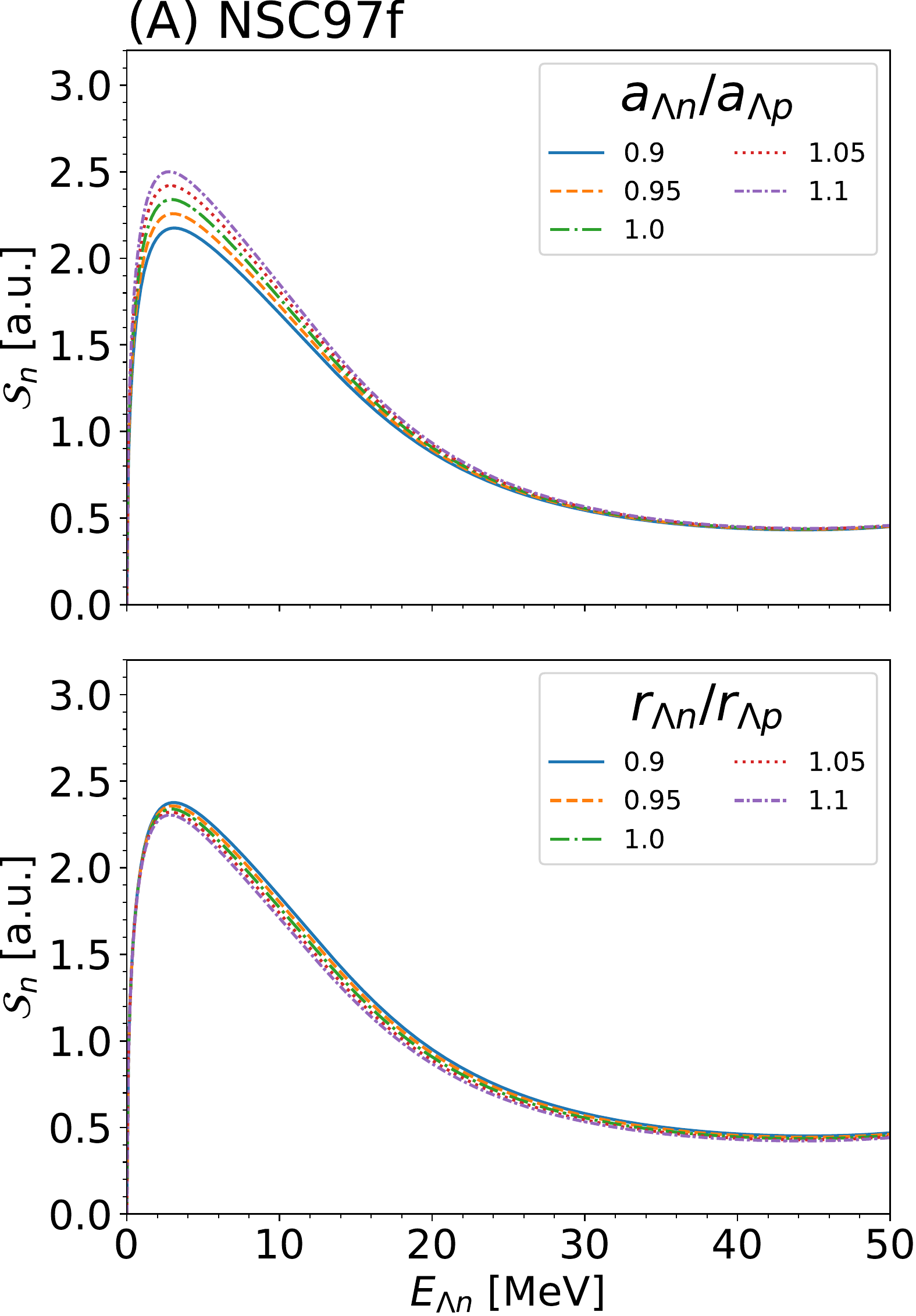}
    % \end{minipage}
    % \begin{minipage}[b]{0.49\linewidth}
    %     \includegraphics[width=4.4cm,clip]{sensitivity_n_B.pdf}
    % \end{minipage}
    % \begin{minipage}[b]{0.49\linewidth}
    %     \includegraphics[width=4.4cm,clip]{sensitivity_r_n.pdf}
    % \end{minipage}
    % \begin{minipage}[b]{0.49\linewidth}
    %     \includegraphics[width=4.4cm,clip]{sensitivity_r_n.pdf}
    % \end{minipage}
    \caption{$\Lambda n$ invariant mass spectra calculated with different $a_{\Lambda n}$ and $r_{\Lambda n}$ values for the $\Lambda n$ process. In the upper plot the value of the scattering length is changed within $\pm 10\%$ of the observed $\Lambda p$ scattering $a_{\Lambda p}=-1.56\fm$, while the value of the effective range varies within $\pm 10\%$ of the observed $\Lambda p$ effective range $r_{\Lambda p}=3.7\fm$ in the lower plot.
        For the two plots $a_{\Sigma N}= 1.68 - i 2.35\fm$ (NSC97f) is used.
        % For the left two plots $a_{\Sigma N}= 1.68 - i 2.35\fm$ (NSC97f) is used, while for the right plots $a_{\Sigma N} = - 3.83 - i 3.01\fm$ (\Julich) is used.
    }
    \label{fig:sensitivity_n_A}
\end{figure}

\begin{figure}[htbp]
    \includegraphics[width=8.6cm,clip]{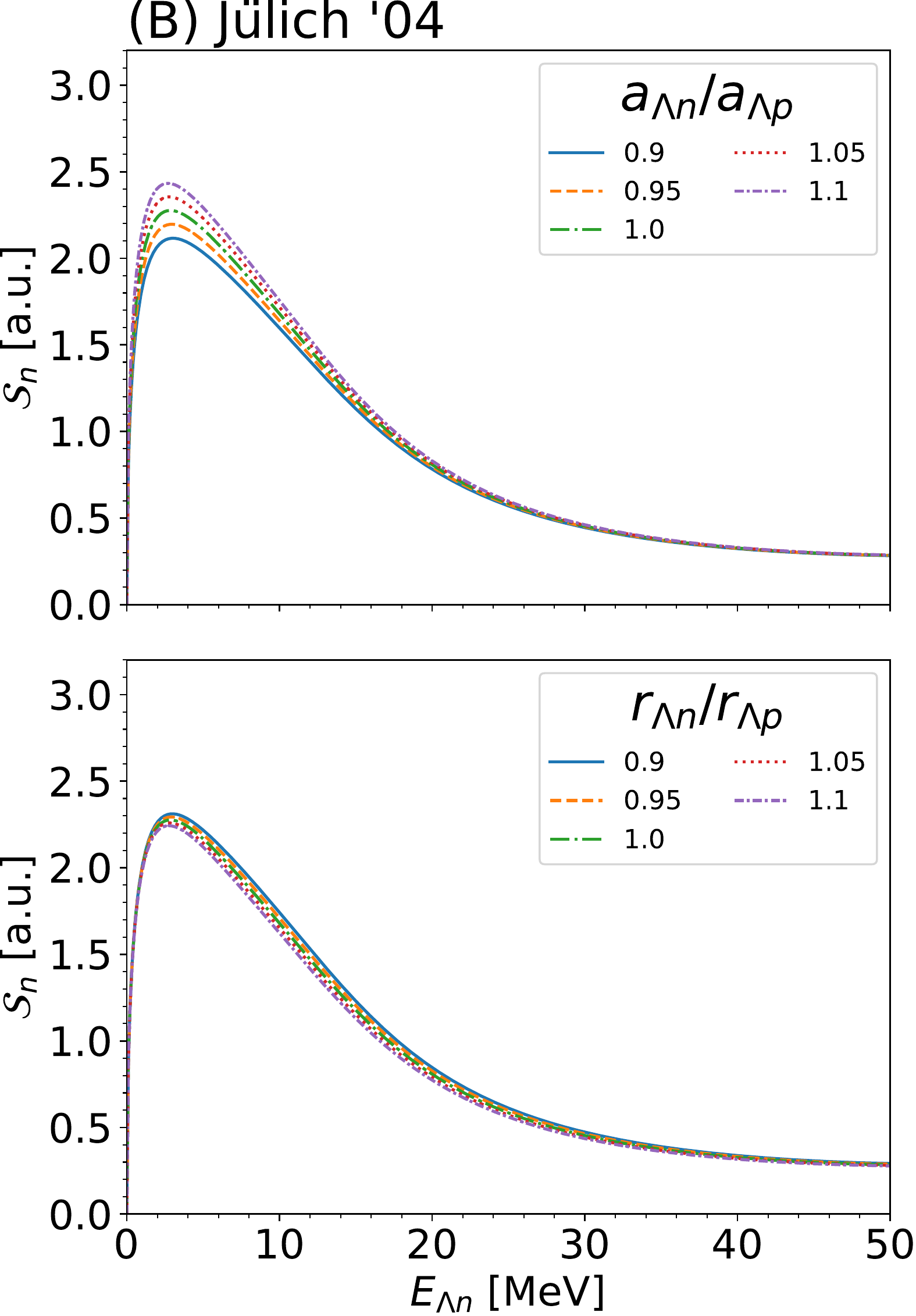}
    \caption{Same as Fig.~\ref{fig:sensitivity_n_A} but $a_{\Sigma N} = - 3.83 - i 3.01\fm$ (\Julich) is used.
        % For the left two plots $a_{\Sigma N}= 1.68 - i 2.35\fm$ (NSC97f) is used, while for the right plots $a_{\Sigma N} = - 3.83 - i 3.01\fm$ (\Julich) is used.
    }
    \label{fig:sensitivity_n_B}
\end{figure}

Next we calculate the invariant mass spectrum for the $\Lambda n$ process. In order to see isospin symmetry breaking, we change the interaction parameters for the $\Lambda n$ process, $a_{\Lambda n}$ and $r_{\Lambda n}$, within within $\pm 10\%$ of the experimentally determined values for $\Lambda p$ scattering. In Fig.~\ref{fig:sensitivity_n_A} and Fig.~\ref{fig:sensitivity_n_B} we show the calculated spectra with different $a_{\Lambda n}$ and $r_{\Lambda n}$ for the $\Lambda n$ process.
In each upper plot we change $a_{\Lambda n}$, while we vary $r_{\Lambda n}$ in each lower plot. One can see from the plots the invariant mass spectra change significantly in the regions $0 < E_{\Lambda n} < 15\MeV$ for the scattering length and less significantly $5 < E_{\Lambda p} < 30 \MeV$ for the effective range. Again we also compare the results with different a values from NSC97f in Fig.~\ref{fig:sensitivity_n_A} and \Julich~ in Fig.~\ref{fig:sensitivity_n_B}.

One can see that the shapes of the spectra with NSC97f and \Julich~are different above $40 \MeV$, while they are almost the same below $40 \MeV$. Thus, we could determine the $\Lambda N$ scattering properties insensitively to the value of $a_{\Sigma N}$ from the invariant mass spectra for $ E_{\Lambda N} < 40\MeV$.

\subsection{Ratio between two reactions}
\label{results_C}
It may be difficult to extract the $\Lambda N$ scattering properties by comparing directly the line shapes of the $\Lambda N$ invariant mass spectra obtained in experiments to that from the theoretical calculation.

Here we would like to propose to take the ratio of the cross sections as a function of the excitation energy $E_{\Lambda N}$ between the $\Lambda n$ and $\Lambda p$ processes:
\begin{align}
    \mathcal{R}_S = 2\frac{{\cal S}_n}{{\cal S}_p}
    \label{eq:r}
\end{align}
where factor 2 is introduced to normalize the ratio to be unity when the isospin symmetry is satisfied. Using $\mathcal{R}_S$,
we expect to
% extract relative difference between the $\Lambda n$ and $\Lambda p$ scattering properties, 
study how different $a_{\Lambda n}$ is from $a_{\Lambda p}$,
that is isospin symmetry breaking.

\begin{figure}[htbp]
    \includegraphics[width=8.6cm,clip]{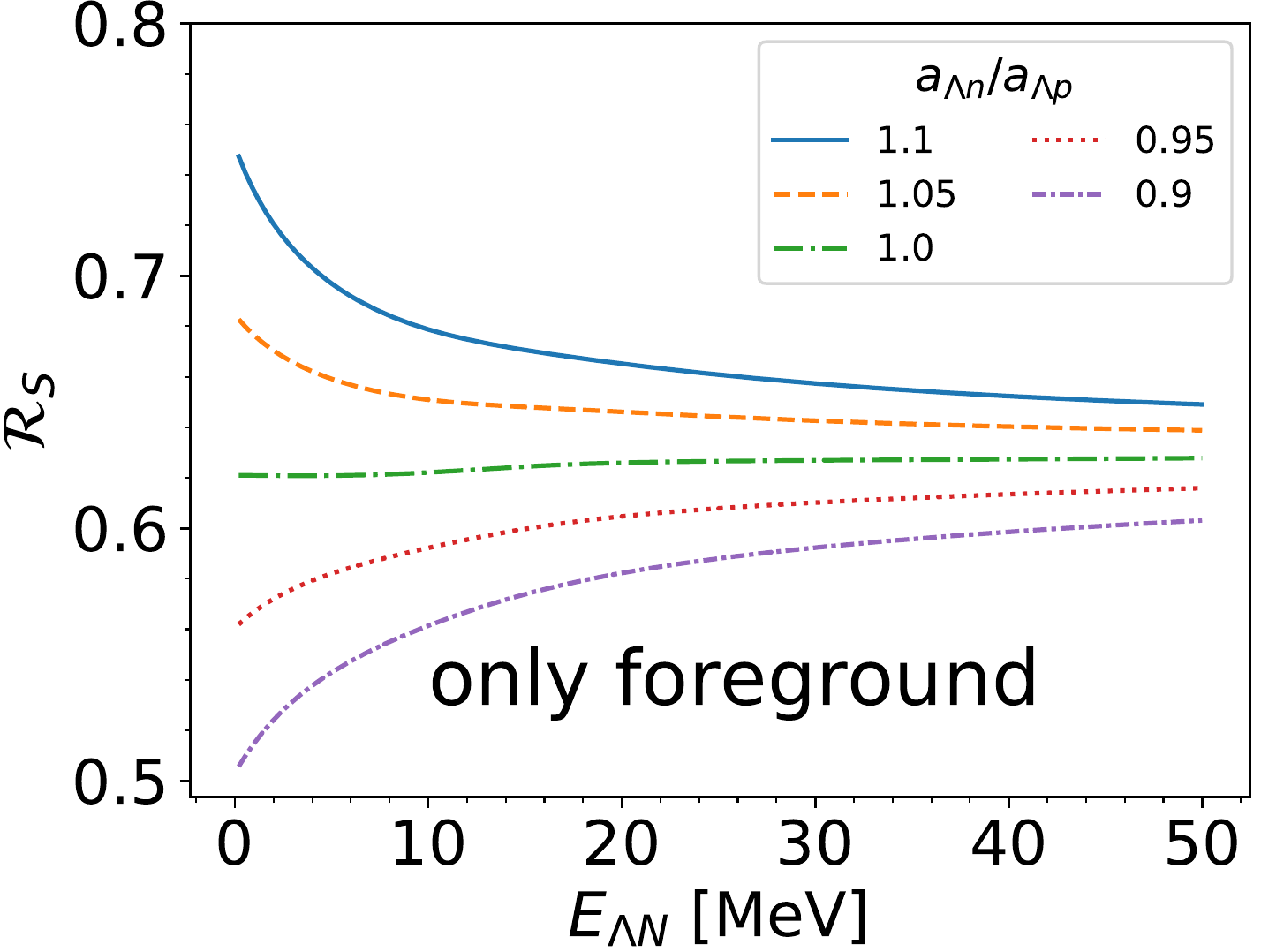}
    \caption{Ratio $\mathcal{R}_S$ calculated only with the foreground diagram as a function of the excited energy $E_{\Lambda N}$.
        We take several $a_{\Lambda n}$ values within $\pm 10\%$ of $a_{\Lambda p}$,
        while the other $\Lambda N$ parameters are fixed at $a_{\Lambda p} = -1.56\fm$, and $r_{\Lambda n} = r_{\Lambda p} = 3.7\fm$.}
    \label{fig:r_FG}
\end{figure}

First of all, we show the ratio calculated only with the foreground diagram in order to check
the feasibility of extracting the isospin symmetry breaking effects in the
$\Lambda N$ interaction from $\mathcal{R}_S$.
In Fig.~\ref{fig:r_FG} we show $\mathcal{R}_S$ for several $a_{\Lambda n}$ values within range of $\pm 10\%$ of $a_{\Lambda p}$.
This range must be much wider than the typical uncertainty in the difference between $a_{\Lambda n}$ and $a_{\Lambda p}$ from isospin symmetery breaking.
% by changing the $\Lambda n$ scattering length within $\pm 10\%$ of the $\Lambda p$ scattering length, which may be wider than a typical range of isospin symmetry breaking. 
We fix the other $\Lambda N$ parameters: $a_{\Lambda p} = -1.56\fm$, $r_{\Lambda n} = r_{\Lambda p} = 3.7\fm$.
Figure~\ref{fig:r_FG} shows that for $a_{\Lambda n}/a_{\Lambda p} < 1.0$ the ratio $\mathcal{R}_S$ tends to go down as the excitation energy approaches to the threshold, which for $a_{\Lambda n}/a_{\Lambda p} > 1.0$ it tends to be enhanced.
Thus isospin symmetry breaking effect on the $\Lambda N$ scattering is clearly seen particularly around the threshold, and we find $\mathcal{R}_S$ to work for studying isospin symmetry breaking in the $\Lambda N$ scattering length.
We see that even though $\mathcal{R}_S$ for the isospin symmetric case $a_{\Lambda n}/a_{\Lambda p} = 1.0$ is almost constant against the excitation energy, it deviates from unity. This is because of the isospin symmetry breaking of the $K^- N \to \pi \Lambda$ amplitudes (the first scattering in Dia.~1), which is independent of the excitation energy.

\begin{figure}[htbp]
    \centering
    \includegraphics[width=8.6cm,clip]{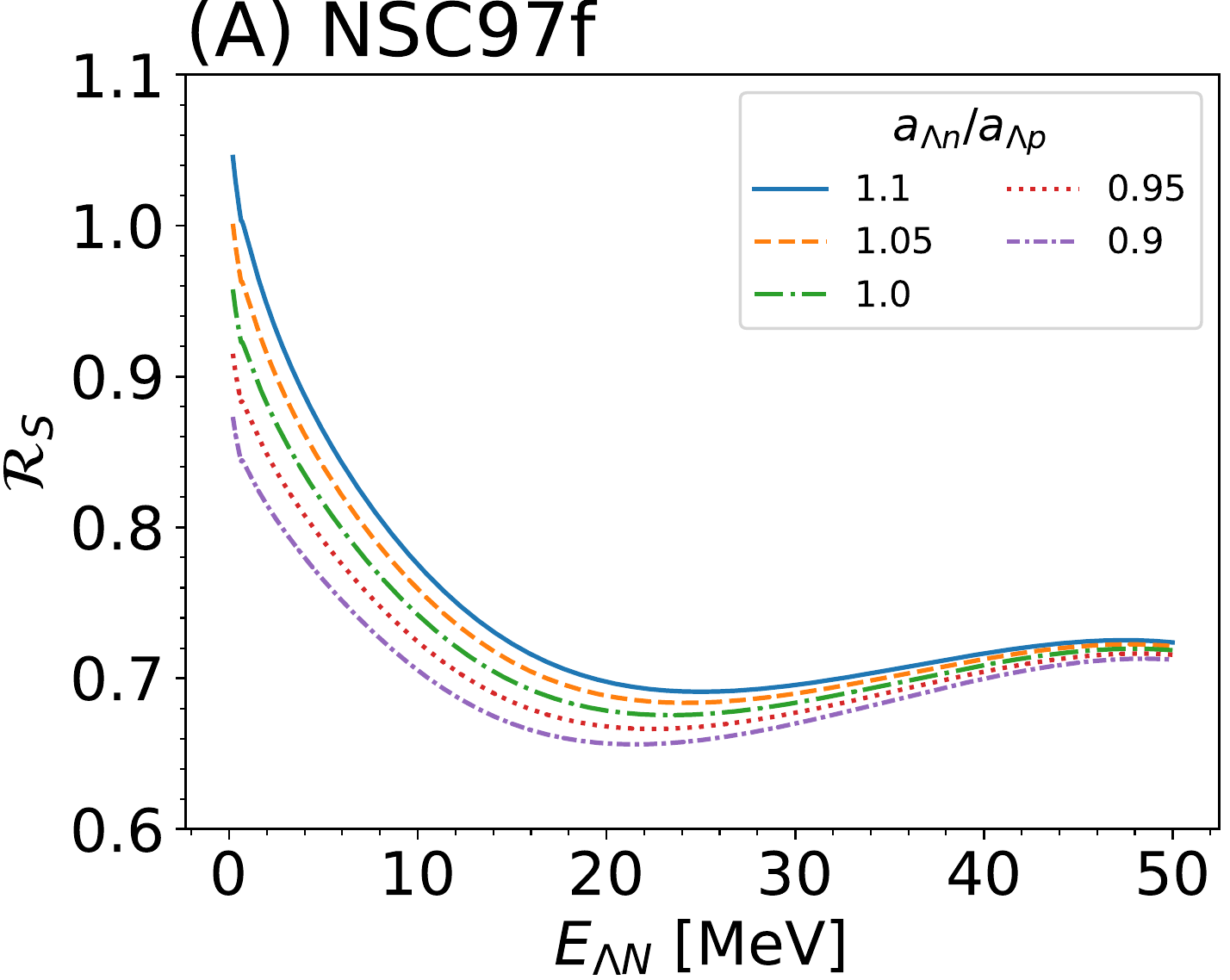}\\
    \vspace{0.3cm}
    \includegraphics[width=8.6cm,clip]{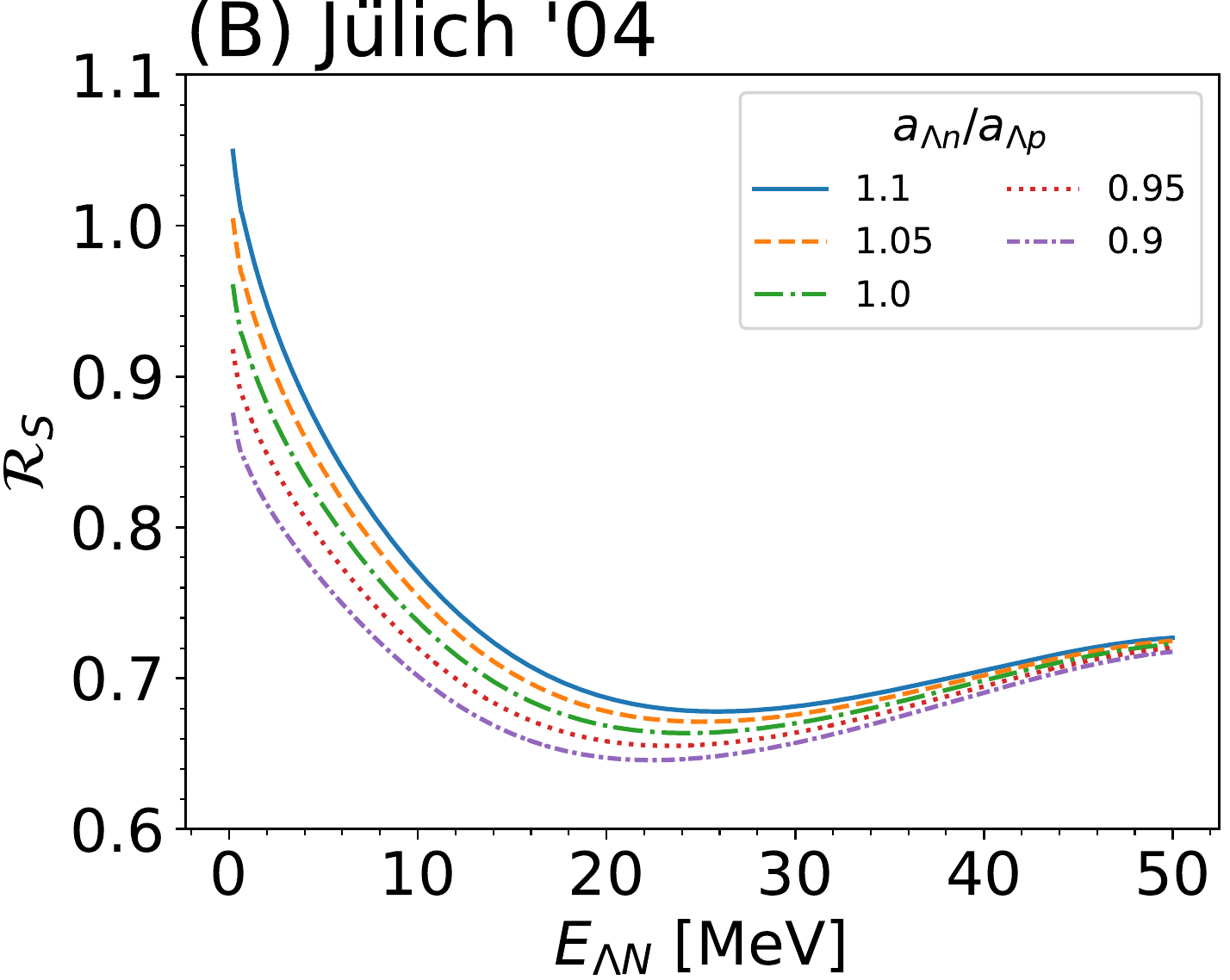}
    \caption{
        Same as Fig.~\ref{fig:r_FG} but for incorporating the background contributions. The upper and lower panels show the ratios obtained by using the $a_{\Sigma N}$ values in NSC97f and \Julich, respectively.
    }
    \label{fig:r_all}
\end{figure}

\begin{figure}[htbp]
    \centering
    \includegraphics[width=8.6cm,clip]{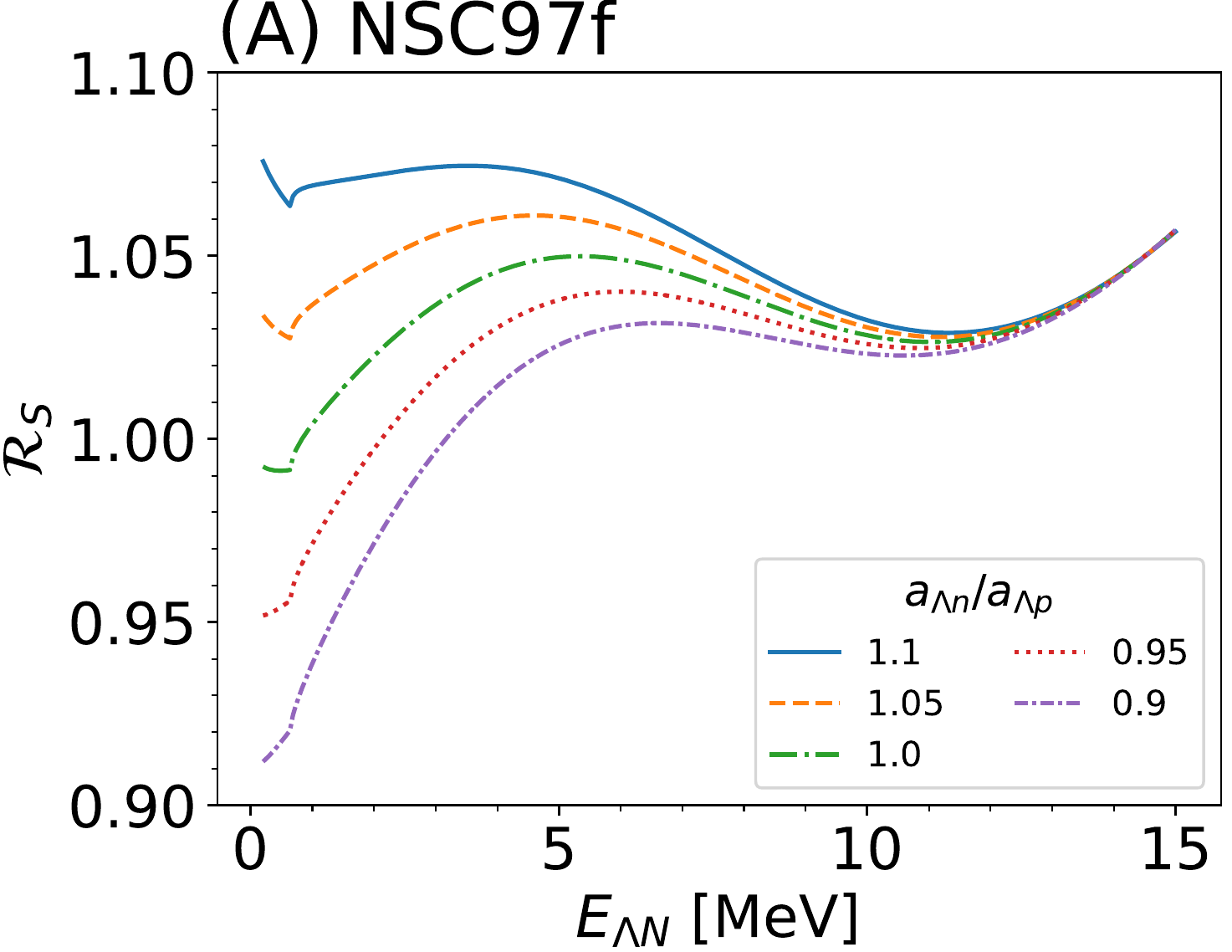}\\
    \vspace{0.3cm}
    \includegraphics[width=8.6cm,clip]{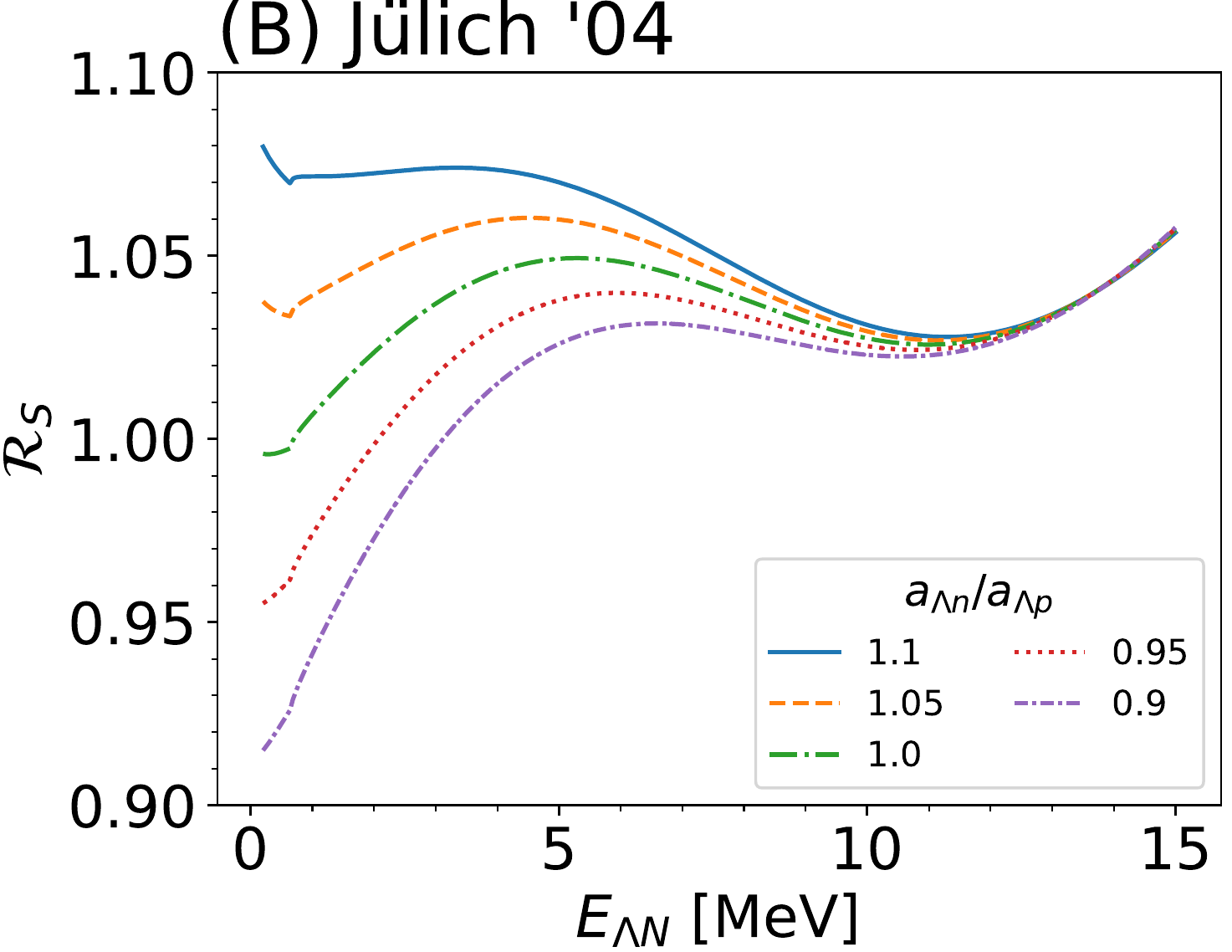}
    \caption{
        Same as Fig.~\ref{fig:r_all} but the cutoff of $\theta_\Lambda^*$ is not applied.
        The upper and lower panels show the ratios obtained by using the $a_{\Sigma N}$ values in NSC97f and \Julich, respectively.
    }
    \label{fig:r_all_pi}
\end{figure}

Next, we calculate the ratio $\mathcal{R}_S$ by incorporating all the background contributions. We change the $\Lambda n$ scattering length $a_{\Lambda n}$ within $\pm 10 \%$ of the $\Lambda p$ scattering length $a_{\Lambda p} = -1.56\fm$ and fix the effective range as $r_{\Lambda n} = r_{\Lambda p} = 3.7\fm$. The results are shown in Fig.~\ref{fig:r_all}. The upper panel (in Fig.~\ref{fig:r_all}) is calculated with $a_{\Sigma N}= 1.68 - i 2.35\fm$ (NSC97f) while the lower panel (Fig.~\ref{fig:r_all}) is with $a_{\Sigma N} = - 3.83 - i 3.01\fm$ (\Julich).
{\red
These figures show that the interference to the background contributions gives rise to an enhancement of $\mathcal{R}_S$ at the vicinity of the threshold.
Still we find qualitative sensitivity to the change of the ratio $a_{\Lambda n}/a_{\Lambda p}$ in a wide range of the excitation energy $0 \le E_{\Lambda p} < 30 \MeV$, in which the ratio $\mathcal{R}_S$ gets enhanced with larger $a_{\Lambda n}/a_{\Lambda p}$. Unfortunately, we do not find such qualitative sensitivity as seen in Fig.~\ref{fig:r_FG}.

Comparing Figs.~\ref{fig:r_FG} and \ref{fig:r_all}, we find that the interference between the foreground and backgrounds is substantially large even if we reduce the background effects by making the angular cut on $\theta_\Lambda^*$. In order to enhance the interference, we calculate the ratio $\mathcal{R}_S$ without the angular cut. The results are shown in Fig.~\ref{fig:r_all_pi}. The difference between $a_{\Lambda n}$ and $a_{\Lambda p}$ can be seen more qualitatively than Fig.~\ref{fig:r_all_pi}. It should be noted that the difference is seen only near threshold up to $E_{\Lambda p} =10 \MeV$. For $a_{\Lambda n}/a_{\Lambda p} < 1.0$, $\mathcal{R}_S$ tends to go down significantly as the excitation energy approaches to the threshold, while for $a_{\Lambda n}/a_{\Lambda p} > 1.0$ it tends not to go down so much. This behavior stems from the effect of the interference between the foreground diagram and the impulse diagram, which is the largest background contribution. This will help us to extract the nature of isospin symmetry breaking on the $\Lambda N$ interaction. At least we could find out the direction of isospin symmetry breaking of the $\Lambda n$ scattering length against $\Lambda p$.

% Let us calculate $\mathcal{R}_S$ in case of no cutoff on $\theta_\Lambda^*$. The results are shown in Fig.~\ref{fig:r_all_pi}.
% The difference between $a_{\Lambda n}$ and $a_{\Lambda p}$ can be seen more clearly than Fig.~\ref{fig:r_FG}.
% For $a_{\Lambda n}/a_{\Lambda p} < 1.0$, $\mathcal{R}_S$ tends to go down significantly as the excitation energy approaches to the threshold, which for $a_{\Lambda n}/a_{\Lambda p} > 1.0$ it tends not to go down so much.
% This is because the effect of the interference between the foreground diagram and the impulse diagram which is the largest before the cutoff on $\theta_\Lambda^*$ is large.
% This will help us to extract the nature of isospin symmetry breaking on the $\Lambda N$ interaction. At least we could find out the direction of isospin symmetry breaking of the $\Lambda n$ scattering length against $\Lambda p$.
}
% It should be noted that, by taking the ratio, we can reduce the model dependence of the $\Sigma N$ scattering length.
% In Fig.~\ref{fig:r_all}, one can see an enhancement of the ratio $\mathcal{R}_S$ around the threshold, $E_{\Lambda N} <~ 1\MeV$ regardless of the value of the $\Lambda n$ scattering length. This is because, as seen in Fig.~\ref{fig:breakdown_2}, around the threshold the contributions of the impulse and kaon exchange diagrams get comparable to that of the foreground. These diagrams contain the $K^- N \to \pi \Lambda$ amplitude and this amplitude has a large isospin breaking at the threshold. This isospin breaking effect surpasses that from the $\Lambda N$ interaction.

\subsection{Comparison to the previous experiments}
\begin{figure}[htbp]
    \centering
    \includegraphics[width=8.6cm,clip]{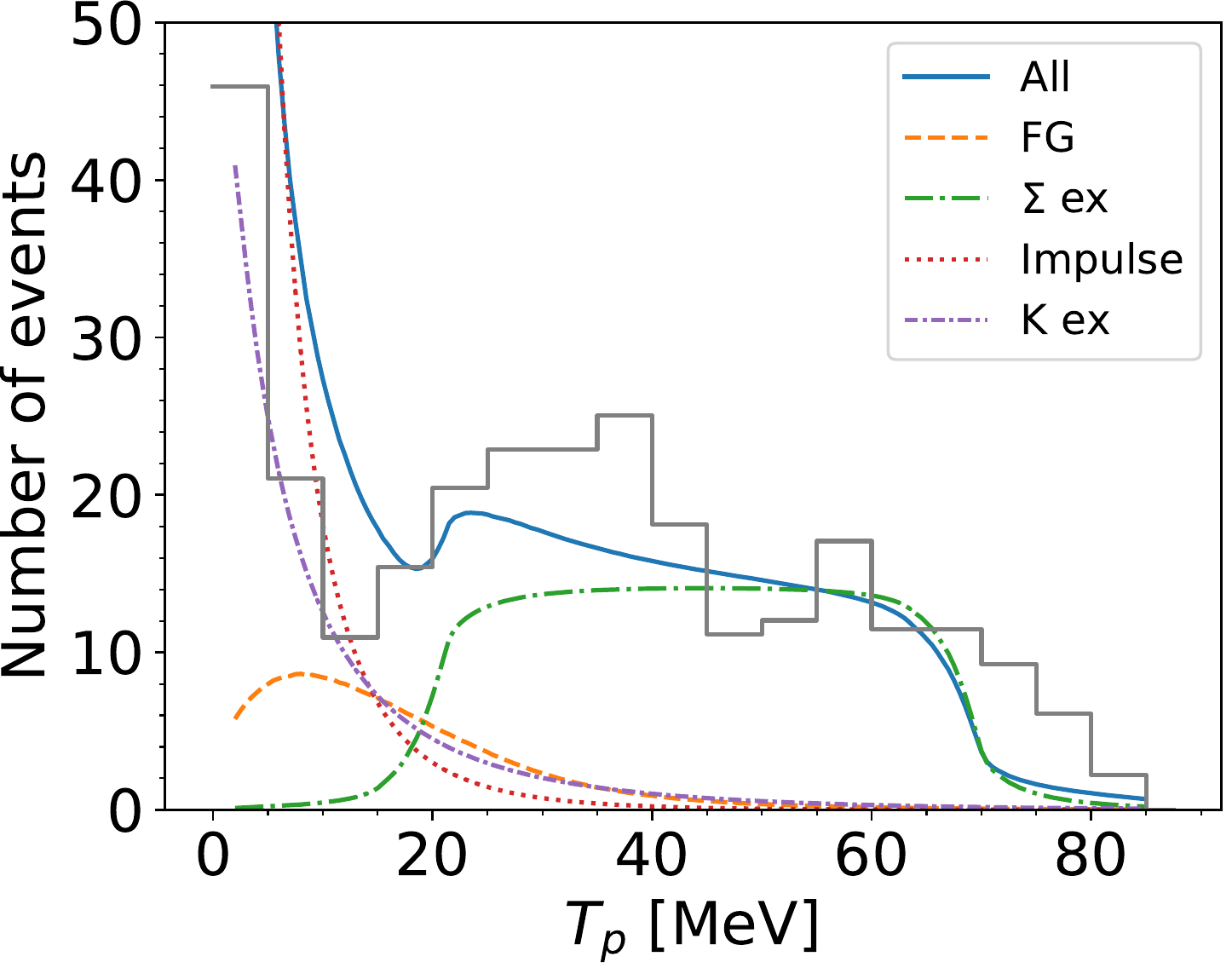}
    \caption{Proton kinetic energy $T_p$ spectrum for $K^- d\to\pi^-\Lambda p$ reaction in comparison with the experimental data taken from Ref.~\cite{Dahl:1961zzb}.
    }
    \label{fig:compare_ex1}
\end{figure}

\begin{figure}[htbp]
    \centering
    \includegraphics[width=8.6cm,clip]{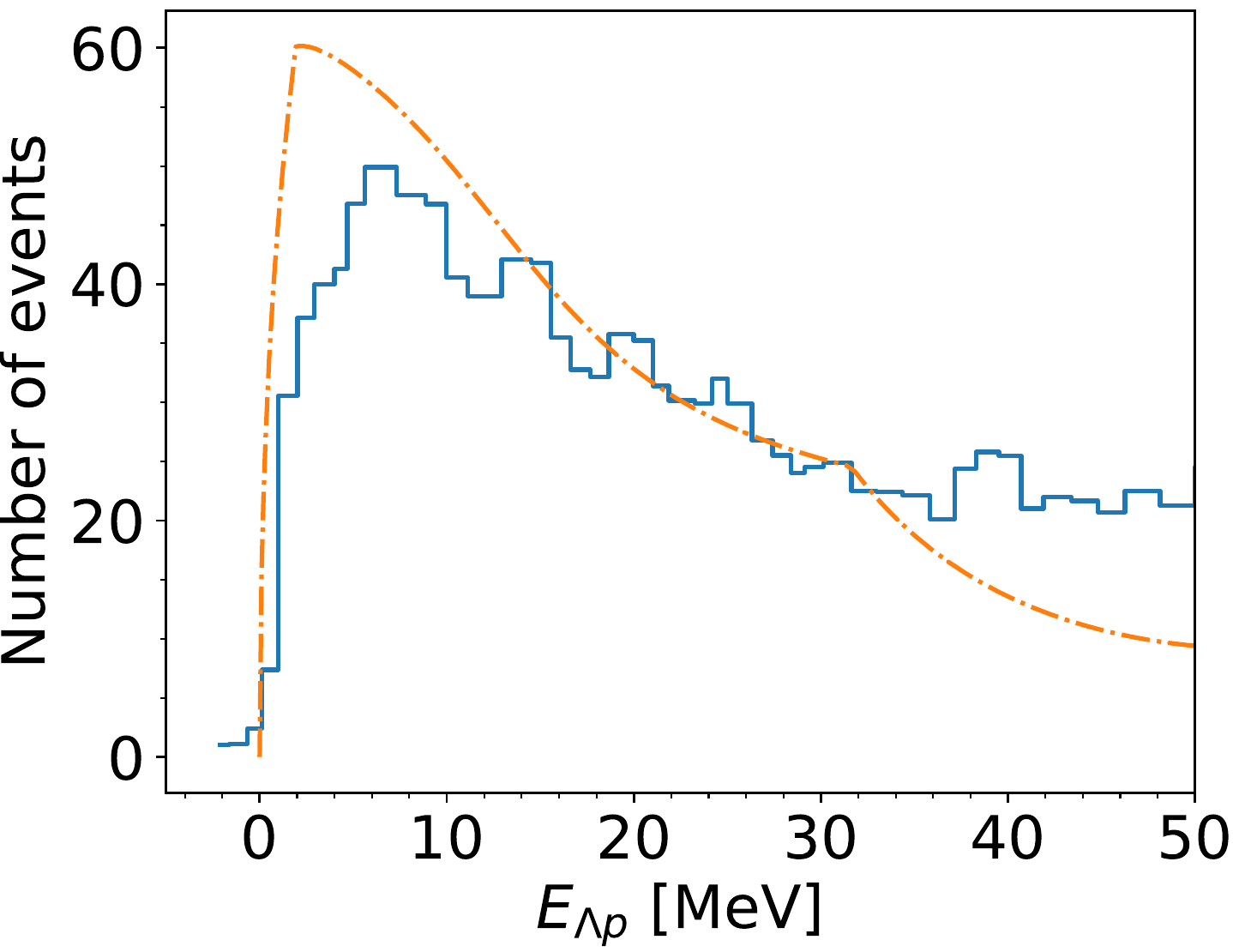}
    \caption{$\Lambda p$ invariant mass spectrum for $K^- d\to\pi^-\Lambda p$ reaction in comparison with the experimental data taken from Ref.~\cite{Tan:1969jq}. The theoretical spectrum is obtained by removing the events with the proton momentum less than $75 \MeV/c$.
    }
    \label{fig:compare_ex2}
\end{figure}

{\red
    We compare our calculation for $K^- d \to \pi^- \Lambda p$ with the past experimental data.

    First we show the proton kinetic energy $T_p$ spectrum in Fig.~\ref{fig:compare_ex1} together with the experimental data in Ref.~\cite{Dahl:1961zzb}. Our calculation is multiplied by a constant to adjust the height to the data. In Ref.~\cite{Dahl:1961zzb}, they have measured the number of counts on the $K^- d \to \pi^- \Lambda p$ reaction with stopped $K^-$.
    In their result, a bump structure was found around $T_p = 30\MeV$, but it was not reproduced in the previous theoretical calculation \cite{Kotani:1959}. Reference \cite{Dahl:1961zzb} mentioned that the bump structure would be explained by the effect of the $\Sigma(1385)$ resonance. Nevertheless, in our calculation, we reproduce the experimental data well without introducing the $\Sigma(1385)$ resonance. In our calculation, we take into account several diagrams with their interference and the bump structure is actually explained by the interference between the $\Sigma$ exchange and other contributions coming from the foreground, the impulse and the kaon exchange diagrams. The theoretical line shown in Ref.~\cite{Dahl:1961zzb} considered only the impulse and $\Sigma$-$\Lambda$ conversion effects without their interference. Our calculation shows that one does not have to introduce the $\Sigma(1385)$ resonance in the calculation such low-energy $\bar K N$ scattering.

    Next we show the $\Lambda p$ invariant mass spectrum in comparison with the experimental data \cite{Tan:1969jq} as plotted in Fig.~\ref{fig:compare_ex2}. In Ref.~\cite{Tan:1969jq}, they have measured the number of counts on the $K^- d \to \pi^- \Lambda p$ reaction with stopped $K^-$ and shown it as a function of the $\Lambda p$ invariant mass. It was pointed out in Refs.~\cite{Dalitz:1982tb,Torres:1986mr} that only events with proton recoil momenta more than $75 \MeV/c$ were counted in Ref.~\cite{Tan:1969jq}. Thus, in order to compare our result with the experimental data given in Ref.~\cite{Tan:1969jq}, we make a similar cut on proton momenta in our calculation. It should be noted that the $\Lambda p$ invariant mass spectrum without the proton momentum cutoff is shown in Fig.~\ref{fig:cross section_all}. Our calculation reproduces well the rapid increase at the threshold seen in the experimental data.
}

\section{Conclusion}
\label{sec:Conclusion}
In this paper, we have studied $K^- d \rightarrow \pi \Lambda N$ reactions with stopped kaons for extracting isospin symmetry breaking in the $\Lambda N$ scattering.
We have proposed that the $K^- d \to \pi \Lambda N$ process has an advantage for the study of isospin symmetry breaking in the $\Lambda N$ scattering, because both isospin partners, $\Lambda p$ and $\Lambda n$, are possible in the final state and we can observe both $\Lambda p$ and $\Lambda n$ final state interactions with the same initial condition. We have formulated the $K^- d \to \pi \Lambda N$ amplitudes by considering not only the foreground contribution which contains the $\Lambda N$ final state interaction but also background contributions which include the impulse diagram, the $\Sigma$, $K$ and $\pi$ exchange diagrams. These background diagrams contains the $\pi\Lambda$ and $\pi N$ final state interactions. For stopped kaons, the $\Lambda N$ interaction is dominated by the spin triplet configuration because of the deuteron spin and $s$-wave dominance of low-energy scattering. In order to reduce the background effects, we have examined the dependence of the amplitude to the angle of $\Lambda$ in the final state and have found that the background effects can be suppressed for narrower angles between $\Lambda$ and $\pi$.

    {\red
        We have found that the $\Lambda N$ invariant mass spectra both for the $\Lambda p$ and $\Lambda n$ processes are sensitive to the $\Lambda N$ scattering properties around the $\Lambda N$ threshold, $E_{\Lambda p} < 30\MeV$ and that one may extract the scattering lengths and the effective ranges from these spectra. It has also turned out that the $\Sigma N \to \Lambda N$ transition effect is less important around the $\Lambda N$ threshold. We have suggested that the ratio of the invariant mass spectra for the $\Lambda n$ and $\Lambda p$ processes works well for the extraction of the ratio of scattering lengths between $\Lambda p$ and $\Lambda n$. We have also compare our calculation with the experimental data for $K^-d\to\pi^- \Lambda p$ reaction and it has been reproduced well.
        % We have seen that the ratio is suppressed around the $\Lambda N$ threshold for $a_{\Lambda n}/a_{\Lambda p} < 1.0$ while it gets enhanced for $a_{\Lambda n}/a_{\Lambda p} > 1.0$.
    }

\begin{acknowledgments}
    The work of Y.I. was partly supported by Grant-in-Aid for Scientific Research from Japan Society for the Promotion of Science (JSPS) (20J20598).
    The work of D.J. was partly supported by Grants-in-Aid for Scientific Research from JSPS (17K05449,21K03530).
    The work of T.I. was partly supported by Grants-in-Aid for Scientific Research from JSPS (19H01902, 19H05141, 19H05181, 21H00114).
\end{acknowledgments}

\appendix
\section{$\Sigma N\to\Lambda N$ transition amplitudes}
\label{appendix_A}
% We show the normalized $\Sigma N\rightarrow\Lambda N$ transition amplitude $f_{\Sigma N\to\Lambda N}$ where $T_{\Sigma N\to\Lambda N}$ is divided by the kinematical factor ${\cal N}$ given in Eq.~\eqref{eq:kf}.
% In this study, we employ the unitarity of $S$-matrix in the isospin-doublet $\Lambda N$ and $\Sigma N$ channels:
In order to obtain the $\Sigma N\rightarrow\Lambda N$ transition amplitude $T_{\Sigma N \to \Lambda N}$, we employ the unitarity of $S$-matrix in the isospin-doublet $\Lambda N$ and $\Sigma N$ channels. The unitarity is implemented to the normalized transition amplitude $f$, which is defined by $T \equiv {\cal N} f$ with the kinematical factor $\cal N$ given in Eq.~\eqref{eq:kf}, as
\begin{align}
    F & = \qty(-V^{-1} + i P )^{-1}
    \label{eq: unitarity}
\end{align}
where $F$, $V$, and $P$ stand for the matrices of the scattering amplitudes, the interaction kernels, and the momenta, respectively, and are defined as
\begin{align}
    F & =
    \begin{pmatrix}
        f_{\Lambda N}      & f_{\Sigma \Lambda} \\
        f_{\Sigma \Lambda} & f_{\Lambda N}
    \end{pmatrix}, \\
    V & =
    \begin{pmatrix}
        v_{11} & v_{12} \\
        v_{12} & v_{22}
    \end{pmatrix}, \\
    P & =
    \begin{pmatrix}
        p^{\, *}_{\Lambda} & 0                 \\
        0                  & p^{\, *}_{\Sigma}
    \end{pmatrix}.
\end{align}
Here, $p^{\, *}_{\Lambda}$ and $p^{\, *}_{\Sigma}$ in $P$ are the momenta of $\Lambda$ and $\Sigma$ in the $\Lambda$-nucleon c.m. frame, respectively. Note that $p^{\, *}_{\Sigma}$ is pure imaginary when one considers the energy region below the $\Sigma N$ threshold. Here we assume that each $v_{ij}$ model parameters is constant.

By using Eq.~\eqref{eq: unitarity}, we obtain the off-diagonal amplitude $f_{\Sigma N\to\Lambda N}$ from the unitarity of $S$-matrix.  The model parameters are determined so as to reproduce the scattering lengths of the $\Lambda N$ and $\Sigma N$ at their thresholds:

\begin{align}
    f_{\Lambda N} & = -a_{\Lambda N},         \\
    f_{\Sigma N}  & = -a_{\Sigma N} = -(A-iB)
\end{align}
with the spin-triplet isospin-doublet $\Lambda N$ scattering length $a_{\Lambda N}$ and $\Sigma N$ scattering length $a_{\Sigma N} = A-iB$ where $A$ and $B$ are real. We obtain the matrix $V$:
\begin{align}
    v_{11} & = \frac{-\kappa_\Sigma B+a_{\Lambda N} \kappa_\Lambda(1-\kappa_\Sigma A)}{\kappa_\Lambda (1-\kappa_\Sigma A+\kappa_\Sigma B a_{\Lambda N} \kappa_\Lambda)}, \\
    v_{12} & =  -\frac{\sqrt{B(1-2\kappa_\Lambda A+\kappa_\Sigma^2 A^2+\kappa_\Sigma^2 B^2)(1+a_{\Lambda N}^2 \kappa_\Lambda^2)}}
    {\sqrt{\kappa_\Lambda} (1-\kappa_\Sigma A+\kappa a_{\Lambda N} B \kappa_\Lambda)},                                                                                   \\
    v_{22} & = \frac{A-\kappa_\Sigma A^2-\kappa_\Sigma B^2+a_{\Lambda N} B \kappa_\Lambda}{1-\kappa_\Sigma A+\kappa_\Sigma a_{\Lambda N} B \kappa_\Lambda}
\end{align}
where $\kappa_\Lambda = p^{\, *}_{\Lambda}$ at the $\Sigma N$ threshold and  $\kappa_\Sigma = -i p^{\, *}_{\Sigma}$ at the $\Lambda N$ threshold.
All the $v_{ij}$ parameters are real. The determined $v_{ij}$ parameters are summarized in Table~\ref{tab:v}. In these calculation, we use the isospin averaged masses to obtain the kinematical variables and find $\kappa_\Lambda = 283.8\MeV$ and $\kappa_\Sigma = 282.6 \MeV$.
\begin{table}
    \caption{Model parameters in a unit of$\MeV^{-1}$}
    \label{tab:v}
    \begin{ruledtabular}
        \begin{tabular*}{8.6cm}{@{\extracolsep{\fill}}ccc}
            $a_{\Sigma N}$ &\begin{tabular}{c}$1.68 - i 2.35\fm$\\ (NSC97f)\end{tabular}      &\begin{tabular}{c}$-3.83 - i 3.01\fm$\\ (\Julich)\end{tabular}  \\ \hline
            $v_{11}$         & $8.3\times 10^{-5}$       & $2.1\times 10^{-2}$    \\
            $v_{12}$         & $6.5\times 10^{-3}$        & $4.4\times 10^{-2}$    \\
            $v_{22}$         & $8.8\times 10^{-3}$      & $7.1\times 10^{-2}$    \\
        \end{tabular*}
    \end{ruledtabular}
\end{table}

\newpage
\bibliographystyle{apsrev4-2}
% \bibliography{references.bib}

\end{document}